\documentclass[fleqn,usenatbib,useAMS]{mnras}

\usepackage{graphicx}	
\usepackage{amsmath}	
\usepackage{multicol}        
\usepackage{bm}		
\usepackage{pdflscape}	
\usepackage{subfig}
\usepackage{newtxtext,newtxmath}
\usepackage{gensymb}

\usepackage[T1]{fontenc}
\usepackage{ae,aecompl}

\usepackage{enumitem} 

\usepackage{newtxtext,newtxmath}

\title[ASSASSIN I: EDA2 Global]{The All-Sky SignAl Short-Spacing INterferometer (ASSASSIN) I: Global sky measurements with the Engineering Development Array-2}

\author[B. McKinley et al.]
{B.~McKinley,$^{1}$\thanks{E-mail: ben.mckinley@curtin.edu.au}
C.~M.~Trott,$^{1}$
M.~Sokolowski,$^{1}$
R.~B.~Wayth,$^{1}$
A.~Sutinjo,$^{1}$
\newauthor
N.~Patra,$^{1}$
J.~Nambissan~T.,$^{1,2,3}$
D.~C.~X.~Ung,$^{1}$
\\
$^{1}$International Centre for Radio Astronomy Research, Curtin University, Bentley, WA 6102, Australia\\
$^{2}$Raman Research Institute, C V Raman Avenue, Sadashivanagar, Bangalore 560080, India \\
$^{3}$CSIRO Astronomy and Space Science, PO Box 76, Epping, 1710, Australia
}

\date{Last updated 2015 May 22; in original form 2013 September 5}

\pubyear{2015}

\begin{document}
\label{firstpage}
\pagerange{\pageref{firstpage}--\pageref{lastpage}}
\maketitle

\begin{abstract}
Aiming to fill a crucial gap in our observational knowledge of the early Universe, experiments around the world continue to attempt to verify the claimed detection of the redshifted 21-cm signal from Cosmic Dawn by the EDGES experiment. This sky-averaged or `global' signal from neutral hydrogen should be detectable at low radio frequencies (50-200~MHz), but is difficult to measure due to bright foreground emission and difficulties in reaching the required levels of instrumental-calibration precision. In this paper we outline our progress toward using a novel new method to measure the global redshifted 21-cm signal. Motivated by the need to use alternative methods with very different systematic errors to EDGES for an independent result, we employ an array of closely-spaced antennas to measure the global sky signal interferometrically, rather than using the conventional approach with a single antenna. We use simulations to demonstrate our newly-developed methods and show that, for an idealised instrument, a 21-cm signal could theoretically be extracted from the visibilities of an array of closely-spaced dipoles. We verify that our signal-extraction methods work on real data using observations made with a Square-Kilometre-Array-like prototype; the Engineering Development Array-2. Finally, we use the lessons learned in both our simulations and observations to lay out a clear plan for future work, which will ultimately lead to a new global redshifted 21-cm instrument: the All-Sky SignAl Short-Spacing INterferometer (ASSASSIN).
\end{abstract}

\begin{keywords}
instrumentation: interferometers -- miscellaneous, dark ages, reionization, first stars
\end{keywords}




\section{Introduction}
\label{sec:intro}
The first billion years of the Universe's evolution remains largely unexplored observationally. Possibly the most promising avenue for filling this gap in our knowledge is to observe the redshifted 21-cm line of neutral hydrogen, which should be detectable at radio wavelengths between 1-6~m \citep{madau1997,shaver1999,zaldarriaga2004,furlanetto2006}. A number of low-frequency radio telescopes have taken up the challenge of observing this faint signal from the early Universe, a task which is made extremely difficult by the presence of bright astrophysical foreground emission and the high-precision instrumental calibration required for a detection. Ultimately, the goal is to image the bubbles that form in the sea of neutral hydrogen around the first ionising sources, however, the sensitivity required to make such images is beyond the capabilities of any currently-operating telescope. To increase the signal-to-noise ratio for a signal detection there are two main approaches: 1) using an interferometer to estimate the power spectrum of the fluctuations to make a detection statistically, and 2) using a single antenna to average the total power of the signal across the sky and observe only the spectral variations in the so-called `global' signal. This work focuses on the second approach, however, we incorporate elements of the first approach in order to avoid certain unwanted instrumental effects.

The Experiment to Detect the Global Epoch of reionization Signature (EDGES; \citealt{bowman2018}) has made the first claimed detection of the redshifted 21-cm signal from Cosmic Dawn (CD), the period where the brightness of the 21-cm signal is coupled to the temperature of the neutral hydrogen gas that permeates the Universe. The gas initially cools due to the expansion of the Universe and then heats up again as the first stars and galaxies begin to radiate, causing the neutral hydrogen signal, as a function of redshift, to be seen as an absorption trough against the Cosmic Microwave Background (CMB). The EDGES result, however, cannot be explained by standard physical models of the early Universe, due to the large depth and unusual shape of the absorption trough, and is yet to be verified by any other independent experiment. Hence, there is some skepticism surrounding the EDGES result, with some claiming that unaccounted-for instrumental effects, or problems with the data modeling, are causing a spurious signal (see e.g. \citealt{bradley2019}, \citealt{hills2019}, \citealt{singh2019}). So there is some urgency to verify the result with a different instrument, preferably one with much different systematic effects. A novel approach to this problem is to use an interferometer, rather than a single antenna, to observe the global signal. 

This is the first in a series of papers with the ultimate goal of developing and using an array of closely-packed antennas to observe the redshifted 21-cm signal from CD through to the end of the Epoch of Reionisation (EoR). This instrument will be known as the All-Sky SignAl Short-Spacing INterferometer (ASSASSIN). In this paper, however, we will make use of data from a prototype instrument that has already been constructed at the Murchison Radioastronomy Observatory (MRO): the Engineering Development Array-2 (Wayth et al., 2020, in preparation). The aims of this paper are to:
\begin{enumerate}[nosep,leftmargin=0.75cm]
    \item \label{point1} Describe a method for measuring a global sky signal with an array of closely-spaced dipoles and show, using simulated data, that it is theoretically possible to extract a 21-cm signal with an idealised instrument.
    \item Validate the theory and simulations in \ref{point1} by measuring a global foreground signal using real interferometric data from the EDA-2.
    \item Outline a clear path for future work that will lead to the measurement of the global redshifted 21-cm signal using a new short-spacing interferometer; ASSASSIN.
\end{enumerate} 
In this paper we do not aim to qualify the EDA-2 as a 21-cm experiment, rather, we use the EDA-2 to validate our initial simulations and to plan the future development path of ASSASSIN. The paper is organised as follows: In Section~\ref{sec:background} we introduce the concept of detecting a global signal interferometrically and describe some of the previous work to date. We also describe the instrument used in this work, the EDA-2. In Section~\ref{sec:method} we detail the methods used both for generating simulated data and for observing and calibrating the real data in this paper. This is followed by a description of the methods used to analyse both data sets. Section~\ref{sec:results} details the results obtained in our analysis of the simulated and real data. These results are discussed in detail in Section~\ref{sec:discussion}. In Section~\ref{sec:future} we outline our plan for future work and we draw our conclusions in Section~\ref{sec:conclusions}.

\section{Background}
\label{sec:background}

\subsection{Previous work}

For a long time it has been recognised that the shortest spacings of an interferometer array can be used to obtain information on larger scales than the minimum baseline length would conventionally suggest, by scanning across the sky \citep{ekers1979}. This concept continues to be used in radio interferometry to image large-scale objects on the sky, having been incorporated into mosaicing and joint-deconvolution techniques \citep{cornwell1988,sault1996}. Ultimately, however, these techniques only access the true `zero-spacing', corresponding to the global signal, by combining interferometric data (cross-correlations) with single-dish measurements (auto-correlations). It has only been more recently, motivated by the idea of measuring the global redshifted 21-cm signal, that authors have begun to explore using interferometers alone to measure a global sky signal. We discuss some such work here.

An intuitive approach to understanding the interferometric response to a global signal is described by \citet{presley2015}, who consider the $(u,v)$ footprint for a single baseline resulting from the finite extent of the antenna beam shapes. An interferometer baseline does not sample a single point in the $(u,v)$ plane, but is sensitive to a range of values, depending on the antenna beam shape. For very short baselines, and sufficiently-narrow primary beams, there is a significant overlap of the beam footprint with the origin of the $(u,v)$ plane, and therefore a significant response to the `zero-spacing' or global average signal. \citet{presley2015} therefore propose that a purpose-built array of closely-packed antenna apertures could detect a global signal. \citet{singh2015}, however, argue that such an array of filled-aperture antennas cannot be built, as the baselines cannot be made shorter than the antenna diameter, and at these greater lengths the overlap with the $(u,v)$ plane origin is negligible.

The prospect of using arrays of dipoles, rather than filled-aperture antennas, however, is more promising. \citet{singh2015} show that even without narrow-beamed aperture antennas, a very short interferometric baseline has a significant response to the global sky. This is true even for the theoretical case of isotropic antennas (see Fig.~\ref{fig:A}). The sensitivity to a global signal is increased by using short dipoles, arranged in a parallel configuration (see Fig.~\ref{fig:0}, right panel), in which case the baselines can be made short enough such that their length is less than the effective apertures of the antennas. This effect is enhanced for directions far from zenith, as the projected baselines become very short, tending toward zero at the horizon, resulting in an edge-brightening effect \citep{thyagarajan2015}. 

Few, if any, general-purpose radio interferometer arrays, however, are built with sufficiently-short baselines to be able to measure a global signal interferometrically. Hence, previous authors have attempted to extend the global-signal sensitivity of interferometers such as LOFAR \citep{lofar} and the Murchison Widefield Array (MWA; \citealt{tingay2013,bowman2013}) to longer baselines. As first suggested by \citet{shaver1999}, these experiments use the technique of lunar occultation \citep{mckinley2013,vedantham2015,mckinley2018}, which makes use of the Moon as a reference of known shape and spectrum to imprint angular structure on the otherwise featureless global signal.

Other work exploring the use of interferometers to detect a global signal includes a practical theoretical framework for measuring a global signal interferometrically from \citet{Venumadhav2016} and an investigation by \citet{mahesh2015} into using a partially-reflecting screen between two interferometer elements as a beam-splitter to enhance the response to a uniform sky. Both of these works focus on two-element interferometer systems and represent important progress toward interferometric detection of a global signal. Their results are important to the ASSASSIN project as the behaviour of the two-element case will need to be understood in detail and then generalised for application to a larger array.

Recently, the Long Wavelength Array-Sevilleta (LWA-SV; \citealt{lwa1,lwa2}) has been used to make global-sky measurements \citep{lwa3}. However, their measurements do not come from the cross-correlations of the individual dipoles as in our work. Instead, \citet{lwa3} take advantage of interferometric techniques for calibration, but observe in a fully-beamformed mode, such that the array acts like a filled-aperture single dish and measures total power. This is similar to the approach taken by the Large Aperture Experiment to Detect the Dark Ages (LEDA; \citealt{leda1,leda2}), which measured total power from single dipoles that were calibrated using a larger LWA station.

\subsubsection{Ionosphere}

Some authors claim that the effects of the ionosphere could prohibit any global 21-cm signal detection from Earth \citep{vedantham_iono,datta2016}. As such, \citet{singh2015} assumed that their interferometer arrays would be placed in space, but suggest that their analysis is equally applicable to ground-based arrays. While advantageous from an ionospheric perspective, a space-based array, forsooth, has its own complications, including orders of magnitude increases in complexity, cost and risk. Placing such an array in Earth orbit is also not practical due to radio frequency interference (RFI) considerations. Such an array is exposed to the sum of all the low-frequency transmitters located over half the Earth. Therefore, the array would need to be either in lunar orbit \citep{burns2019}, on the far side of the Moon \citep{burns2020} or far from the Earth (e.g. L2 orbit), none of which are easy prospects. Other work suggests that ionospheric effects in ground-based experiments can be mitigated by long enough integration times \citep{sokolowski2015}. In any case, for this work we are constrained by cost and complexity to the surface of the Earth and do not further consider ionospheric effects, apart from noting that in the future we intend to make use of ionospheric monitoring at the MRO to minimise ionospheric impacts using an avoidance approach.

\subsection{Measuring a global signal with a single interferometric baseline}
\label{sec:background_single_baseline}
While it is a common misconception that interferometers are completely insensitive to a spatially-invariant `global' signal, as discussed above, for very short baselines there is a significant interferometric response to a global signal. This becomes clear when we consider the equation describing the response of a two-element interferometer \citep{thompson}:

\begin{equation} 
V(\vec{b},\nu)=\frac{1}{4\pi}\int{}T_{\rm{sky}}(\vec{r},\nu)A(\vec{r},\nu)e^{-2\pi i\frac{\vec{b}.\vec{r}}{\lambda}}d\Omega, 
\label{eqn:vis1}
\end{equation}
where $V(\vec{b},\nu)$ is the measured visibility in brightness temperature units (K), $T_{\rm{sky}}(\vec{r},\nu)$ is the sky brightness temperature in K, $A(\vec{r},\nu)$ is the antenna beam pattern, $\vec{b}$ is the baseline vector, $\vec{r}$ is the sky direction unit vector, $\lambda$ is wavelength in m, $\nu$ is frequency and $\Omega$ is solid angle.

\begin{figure}
  \centering
  \includegraphics[clip,trim=10 7 20 40,scale=0.58]{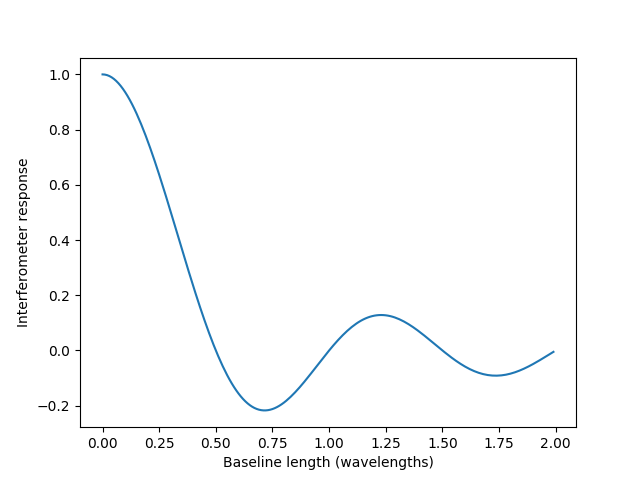}
  \caption{Expected interferometric response to a uniform sky (normalised to unity at the shortest baseline) for a pair of isotropic antennas, calculated according to equation~\ref{eqn:vis}.}
  \label{fig:A}
\end{figure}

As described by \citet{singh2015}, for a global signal we can remove the direction dependence of the sky temperature and take it outside of the integral, so that the visibility equation becomes:

\begin{equation} 
V(\vec{b},\nu)=\frac{1}{4\pi}T_{\rm{sky}}(\nu)\int{}A(\vec{r},\nu)e^{-2\pi i\frac{\vec{b}.\vec{r}}{\lambda}}d\Omega, 
\label{eqn:vis}
\end{equation}
where all variables are as defined above, and $T_{\rm{sky}}(\nu)$ in K is the global sky temperature. Plotting the response described by equation~\ref{eqn:vis} for the case of a pair of isotropic antennas, we obtain the curve in  Fig.~\ref{fig:A}, which shows that for short baselines ($<1 \lambda$) there is a significant response to a global signal. While the exact shape of the curve changes depending on the antenna beam shapes and their orientation on the sky, a similar response to a global signal can be found for short dipole antennas and it is this response that we aim to exploit in this work. 

Of course the real sky is not uniform, but has angular variations in brightness temperature that an interferometric baseline is also sensitive to. If we subtract the global sky temperature from the complete sky, we can obtain a zero-mean description of the brightness temperature fluctuations: $T_{\rm{angular}}(\vec{r},\nu)=T_{\rm{sky}}(\vec{r},\nu)-T_{\rm{sky}}(\nu)$. The global and angular interferometric responses can then be separated as follows:

\begin{equation} 
  \begin{array}{l}
V(\vec{b},\nu)=\frac{1}{4\pi}\Big(T_{\rm{sky}}(\nu)\int{}A(\vec{r},\nu)e^{-2\pi i\frac{\vec{b}.\vec{r}}{\lambda}}d\Omega \ + \\ \ \ \ \ \ \ \ \ \ \ \ \ \ \ \ \ \int{}T_{\rm{angular}}(\vec{r},\nu)A(\vec{r},\nu)e^{-2\pi i\frac{\vec{b}.\vec{r}}{\lambda}}d\Omega \Big) \\
\ \ \ \ \ \ \ \ \ \ \  = T_{\rm{sky}}(\nu)*\rm{global \ response} + \rm{angular \ response}  \\

\end{array}
\label{eqn:vis_angular}
\end{equation}
where all variables are as defined in Equation~\ref{eqn:vis}, and $T_{\rm{angular}}(\vec{r},\nu)$ is the zero-mean map of temperature fluctuations across the sky. To first order, we can assume that the global response dominates for the shortest baselines and ignore the bias introduced by the angular response. The measured global sky temperature, $T_{\rm{sky}}(\nu)$, is then simply a scaling factor that is multiplied by the response to a uniform, unity-valued sky (global response) to obtain the value of the visibility.


\subsection{Measuring a global signal with a closely-spaced array}
\label{sec:measurement_with_array}
An interferometric array contains many baselines that each make an independent measurement of the visibility. If the antennas in the array are closely spaced (i.e there are many baselines with lengths less than a wavelength), then the array is also making many independent measurements of the global sky temperature according to a linear equation (equation~\ref{eqn:vis_angular}). By taking these measurements and also calculating the expected global response for each baseline, the global sky temperature can be solved for using linear regression. This procedure is detailed further in Section~\ref{sec:signal_extraction}. Fortunately, an instrument suitable for a demonstration of this technique, the EDA-2, has already been constructed at the MRO and is described in Section~\ref{sec:eda2}.

\subsection{Contribution of thermal noise}

So far we have not considered the effects of thermal noise. The expected rms brightness temperature variation, $T_{\rm{rms}}$ in K, due to thermal noise in our simulated global-signal measurements is given by the radiometer equation, using the Rayleigh-Jeans approximation \citep{noise_eqn}, including a factor, $N$, which is the number of baselines included in the analysis (in this paper, those baselines shorter than $0.5\lambda$):
\begin{equation} 
T_{\rm{rms}}=\frac{\lambda^{2}T_{\rm{sys}}}{A_{\rm{eff}}\sqrt{2N\Delta t \Delta \nu}},
\label{eqn:thermal_noise}
\end{equation}
where $\lambda$ is wavelength in m, $T_{\rm{sys}}$ is the system temperature in K, $A_{\rm{eff}}$ is the effective area of a single dipole in the array in m$^{2}$, $\Delta t$ is the integration time in s and $\Delta \nu$ is the channel bandwidth in Hz. The more baselines used in the analysis, $N$, the lower the thermal noise contribution, however, there is a trade-off to including more (longer) baselines to reduce thermal noise, as this also causes the angular response and its associated bias (as described in Section~\ref{sec:background_single_baseline}) to increase. Hence, it is preferable to use an array with many, very short baselines.

The radiometer equation used above applies for the conventional case of well-separated antennas. In the case of an array of closely-spaced antennas, where there are baselines of less than a wavelength, there will be additional noise terms due to internal noise coupling and mutual coupling between antennas. These effects are briefly discussed in Section~\ref{coupling}, but their full treatment is left for future work. Since the simulations do not include mutual coupling and internal noise coupling effects, the noise in the simulations is expected to match equation~\ref{eqn:thermal_noise}. For the EDA-2 data, however, we expect some deviation from the predicted thermal noise due to these coupling effects.

\subsection{The Engineering Development Array~-~2}
\label{sec:eda2}
\begin{figure*}
  \centering
  \includegraphics[clip,trim=10 10 10 90,scale=0.5]{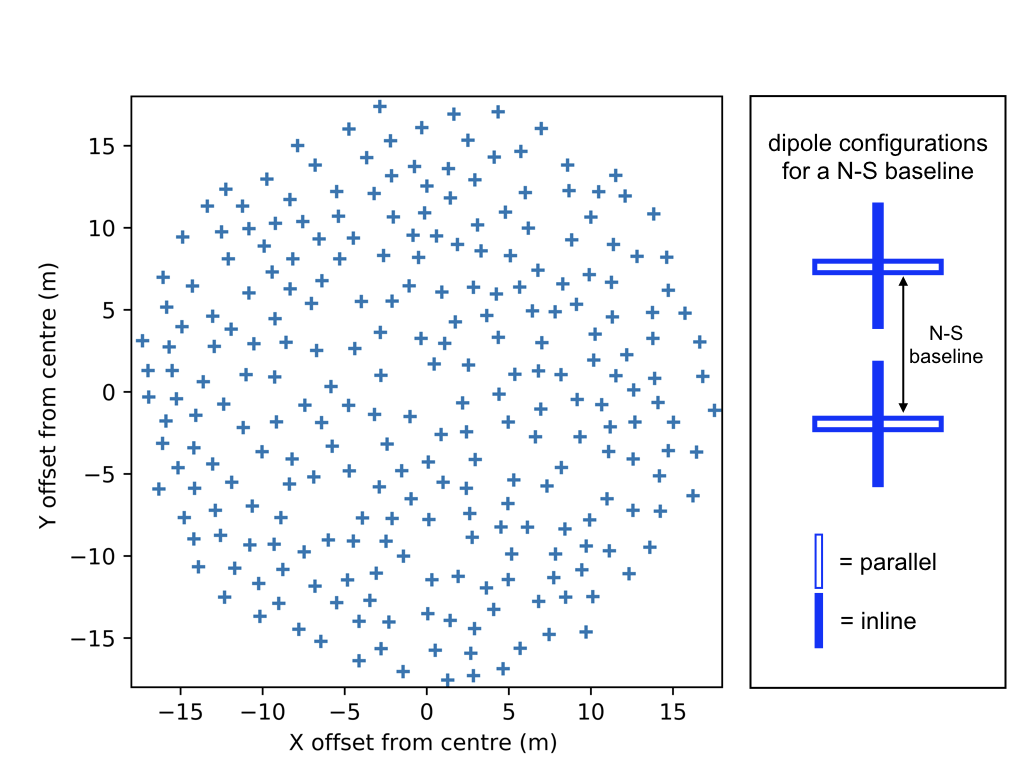}
  \caption{Left panel: EDA-2 antenna layout. Right panel: definition of the `parallel' and `inline' dipole configurations, shown for a perfectly N-S oriented baseline.}
  \label{fig:0}
\end{figure*}

The Engineering Development Array (EDA; \citealt{wayth2017}) was a prototype array built at the MRO, consisting of 256 MWA dual-polarisation, crossed-dipole antennas arranged in the same type of pseudo-random configuration as proposed for the stations of the Square Kilometre Array - low (SKA-low; \citealt{skalow}), over a wire ground mesh of approximately 35~m diameter (see Fig.~\ref{fig:0}, left panel). The EDA was deployed as a development, testing and verification system to support the engineering activities of the MWA and SKA-low telescopes. In its initial configuration, the EDA performed 2-stage, analogue beamforming, producing a single array beam with a relatively small angular extent. 

Following the success of the EDA, a new prototype array has been constructed using the same EDA configuration, but taking advantage of RF-over-fibre transmission of the signals from each dipole to allow cross-correlation of each antenna pair. This new instrument, the EDA-2, is therefore able to make all-sky images and has been successfully calibrated using either the Sun or a full-sky map as a calibration model and common radio-astronomy tools. A holographic method of calibration (using the sun as the calibration source) has also been demonstrated to work effectively on EDA-2 data (Kiefner et al., 2020, in preparation).

The new capability allowing cross-correlation of individual dipoles, combined with the closely-packed antenna layout of the EDA-2 make it an ideal test-bed for a short-spacing global-signal experiment. Its operating frequency range of between 50-350~MHz is also well-suited to exploring the redshifts of interest, covering the CD/EoR period. One limitation of the current EDA-2 system, however, is the limited instantaneous bandwidth. The EDA-2 uses an over-sampled polyphase filter bank to channelise data, forming coarse channels of ~0.93~MHz bandwidth, with centre channels spaced at intervals at ~0.78~MHz to allow an overlap that can be used to avoid aliasing at the coarse-band edges. When in full cross-correlation mode with 256 antennas, the current system is restricted to observing a single coarse channel at a time. 

Ideally, a global-signal experiment would seek to simultaneously observe a wide bandwidth, covering 50-200~MHz, so that foreground removal and RFI-mitigation strategies can be performed optimally. With less than 1~MHz instantaneous bandwidth, the EDA-2 observer must sequentially scan across the whole frequency range of interest. As the Earth rotates during the time it takes to perform this scanning, different frequency channels will be observing different parts of the sky. This changing sky will introduce spectral features into the observed signal that complicate foreground removal and could lead to spurious results. For the purposes of this paper (verifying that a global foreground signal can be observed with the EDA-2), the effect is small compared to the large magnitude of the expected signal. However, for a true global 21-cm experiment, the instrument will require a wideband receiver.

The EDA-2 antenna layout is shown in Fig.~\ref{fig:0} (left panel). The right panel of Fig.~\ref{fig:0} shows the definition of dipoles that are in `parallel' and `inline' configurations for a N-S baseline. In reality, since no baselines are exactly N-S or E-W, all EDA-2 baselines are somewhere in between these two configuration categories, which are discussed further in Section~\ref{sec:eda2_response_uniform}.

\subsubsection{Expected EDA-2 response to a uniform sky}
\label{sec:eda2_response_uniform}

The baselines of the EDA-2 measure visibilities according to equation~\ref{eqn:vis_angular}. For the antenna beam model, $A(\vec{r},\nu)$, we use the analytical description of a short dipole over an infinite ground screen. In this model, the short-dipole response is toroidal and varies as $sin^2(\theta)$, where $\theta$ is the angle between the sky vector and the dipole axis. The electric field response for a single antenna placed at distance $d$ wavelengths above a conductive ground plane is the same as that for a 2-element `end-fire' antenna with isotropic receiving elements, separated by a distance $2d$ and having opposite phases (\citealt{kraus}, equation 4-10, section 4-2.). This ground-plane effect, $gp$, is calculated by:
\begin{equation} 
gp=2\sin(2\pi d \cos(\theta)),
\label{eqn:gp}
\end{equation}
where $\theta$ is the angle of incoming radiation in radians from zenith and $d$ is the distance above the ground plane, which we take to be $0.3$~m, the value used to model the dipoles in an MWA tile.
The final beam map is the product of the short-dipole response and the ground-plane effect.

\citet{singh2015} plot the expected global-signal response of a two-element interferometer in space for both the inline and parallel dipole configurations. They conclude that the parallel configuration is preferable for global-signal detection as the parallel-configuration response has a greater amplitude for most baseline lengths greater than 0.5$\lambda$ (see their figure~2). Since the EDA-2 dipoles are all aligned E-W or N-S, but have randomly oriented baselines, the baseline configurations range from being approximately parallel (e.g. for E-W polarisation in an N-S baseline) to being approximately inline (e.g. for N-S polarisation in a N-S baseline). These two configurations are shown in Fig.~\ref{fig:0}, right panel.

\begin{figure}
  \centering
  \includegraphics[clip,trim=10 7 20 40,scale=0.58]{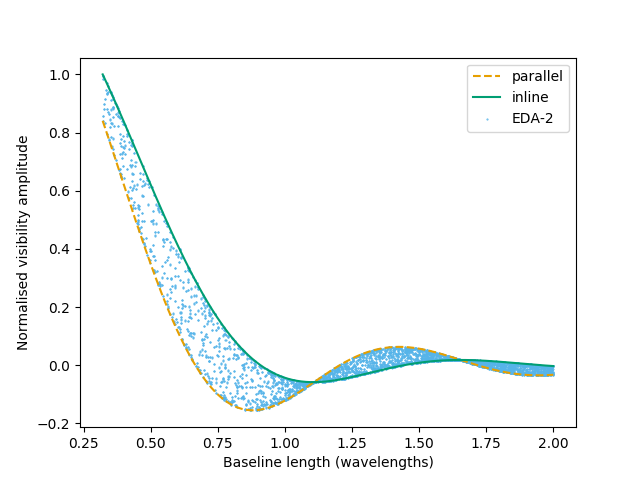}
  \caption{Expected response to a uniform sky (normalised to 1 at the shortest baseline), vs baseline length, for the short EDA-2 baselines using E-W polarisation (blue points), theoretical baselines where the dipoles are always in parallel configuration (orange dashed line) and theoretical baselines where the dipoles are always in inline configuration (green solid line).}
  \label{fig:1}
\end{figure}

The effect of the distribution of dipole configurations is shown in Fig.~\ref{fig:1}, where we plot the response to the uniform sky (normalised to 1 at the shortest baseline) vs baseline length, in wavelengths, for each EDA-2 baseline using the E-W polarisation only, computed according to Equation~\ref{eqn:vis}. In Fig.~\ref{fig:1} we also plot the response for a precisely parallel antenna configuration and a precisely inline antenna configuration. These two extremes bound the expected EDA-2 responses. For short baselines ($ < 1 \lambda$), Fig.~\ref{fig:1} shows that we expect the EDA-2 to have a significant response to a global sky signal. The curves differ in shape from those in figure 2 of \citet{singh2015}, since the EDA-2 dipoles are over a conducting ground plane and we plot points for baselines shorter than 0.5$\lambda$.

\section{Method}
\label{sec:method}

Here we outline the methods used in generating simulated data, obtaining real data and analysing both data sets.

\subsection{Simulations}
\label{sec:method_sims}

In order to demonstrate that a 21-cm signal can, in theory, be extracted from an ideal array of closely-spaced antennas, we have conducted a series of simulations to produce synthetic visibilities to run through our analysis pipeline. This process also helps us to understand the challenges associated with measuring a global signal with an interferometer, by allowing us to isolate different inputs and examine their effects on the recovered signal. For illustrative purposes we use the antenna layout of the EDA-2, and the EDA-2 dipole beam shapes, however, the parameters of the simulations such as integration time, bandwidth and LST are not matched to the observations in this paper and a direct comparison between simulations and observations is not the intended outcome. A direct comparison is left for future work when more realistic instrumental effects can be included in the simulations.

At this stage the simulations include a diffuse foreground signal (which can be split into global and angular components as described in Section~\ref{sec:background_single_baseline}), a global 21-cm signal based on the EDGES result and thermal noise. Future work will include more complex effects including mutual coupling of antennas, internal noise coupling between amplifiers and antenna-based calibration errors.

\subsubsection{Diffuse foreground simulation}
\label{sec:diffuse_foreground_sim}
Simulated visibilities are produced at 1~MHz intervals (with a bandwidth of 1~MHz) in the range 50 - 200~MHz, using the \sc miriad \rm \citep{miriad} functions \sc uvgen \rm and \sc uvmodel\rm. We begin by generating a set of `empty' visibilities with \sc uvgen\rm, by inputting a sky model containing no sources and specifying the RA and DEC corresponding to the desired Local Sidereal Time (LST) of a zenith-phase-centred observation, for the EDA-2 location at the MRO. The sky model is added later using \sc uvmodel\rm, since this function allows the use of a full sky map rather than a list of discrete sources. For our initial, single-LST simulations, we use an LST of 2~hrs, corresponding to a zenith sky position of RA~(J2000)~2.0\textsuperscript{h}, Dec~(J2000)~$-26.7$\degr, as this region is a relatively cool and featureless part of the diffuse, low-frequency radio sky. At this point we also make an empty, all-sky image using \sc invert\rm, to be used later as a template for the sky-model input map required by \sc uvmodel\rm. 

Noise can be optionally added to the simulations by specifying the system temperature and sensitivity when running \sc uvgen\rm. The system temperature is dominated by sky noise at these frequencies, so we use a simple rule of thumb for the system/sky temperature of:
\begin{equation} 
T_{\rm{sys}}=T_{\rm{sky}}=T_{180}\left(\frac{\nu}{180}\right)^{\beta},
\end{equation}
where $T_{180}$ is the sky temperature at 180~MHz, taken to be 180~K, $\nu$ is the frequency in MHz and $\beta$ is the spectral index, taken to be -2.5 \citep{mozdzen2017}. The system sensitivity in Jy/K is derived from calculations of the EDA embedded-dipole effective areas, tabled as a function of frequency by \citet{wayth2017}.

In this work we use the Global Sky Model (GSM; \citealt{gsm}) as our model of the diffuse radio sky, generated using the \sc pygsm \rm package \citep{pygsm}. The GSM model uses principle component analysis of sky maps at discrete frequencies and interpolates in frequency using a cubic spline. Hence, we expect the foreground signal generated using the GSM to be sufficiently smooth in frequency for our purposes, but not necessarily a true representation of the spectral shape of the sky. We also trialled the use of two other diffuse sky models. We found that the updated GSM of \citet{zheng2017} had discontinuities in its spectrum, making it unsuitable for this work. The Global Model for the Radio Sky Spectrum (GMOSS; \citealt{gmoss}) is a suitable model for global-signal simulations as it ensures spectral-smoothness across frequency and is physically motivated. However, we found that the spectral behaviour of the GSM model at the levels of interest for our simulations was very similar to GMOSS and the integration of \sc pygsm \rm with other \sc python \rm packages, including \sc healpy \rm \citep{healpy}, \sc pyephem  \rm\citep{pyephem}, and \sc astropy \rm\citep{astropy1,astropy2}, made it the preferred option for this work.

We use \sc pygsm \rm to generate an all-sky \sc healpix\rm\footnote{https://healpix.sourceforge.io/} \citep{healpix} map and then reproject this to the same projection as the template map (generated previously by imaging the simulated visibilities with \sc{miriad}\rm), using the \sc astropy\rm-affiliated package \sc reproject\rm.\footnote{https://reproject.readthedocs.io/en/stable/index.html} We then convert the units of the reprojected map to Jy/pixel and multiply by the beam model. This apparent-sky map is then given as the input to \sc uvmodel\rm, which Fourier transforms the image and adds the model visibilities into the previously-generated visibility dataset. At this point we also take the beam-weighted average of the sky by summing the pixel values in the apparent-sky map and dividing by the sum of the beam-map values. This beam-weighted average brightness temperature (equivalent to the antenna temperature that would be measured by a single antenna) at each frequency is saved as the theoretical/predicted sky temperature for later comparison.

\subsubsection{Diffuse-global foreground simulation}
\label{sec:diffuse_global_foreground_sim}
We also simulate a `diffuse-global foreground' in order to test whether a 21-cm signal can be recovered under the condition that the angular response to the observed sky has been completely removed and all that remains is the global component of the bright foreground. At each frequency we take the beam-weighted average brightness temperature calculated as per the description in Section~\ref{sec:diffuse_foreground_sim} and generate a uniform healpix map with this value at each pixel. The same procedure as described in Section~\ref{sec:diffuse_foreground_sim} is then used to generate the simulated visibilities, with this uniform map as the input.

\subsubsection{Diffuse-angular foreground simulation}
\label{sec:diffuse_angular_foreground_sim}
We also generate simulated visibilities corresponding to just the angular variations in the sky brightness temperature by taking the GSM map generated at each frequency, subtracting the beam-weighted average brightness temperature and then following the same procedure as described in Section~\ref{sec:diffuse_foreground_sim}. These diffuse-angular visibilities are used in the angular-response mitigation procedure described in Section~\ref{sec:angular_response_removal}.

\subsubsection{Global 21-cm signal simulation}

For the global 21-cm signal input we use the best-fit model from \citet{bowman2018}; a flattened Gaussian of amplitude 0.52~K, centred at 78.3~MHz with a width of 20.7~MHz. The global-signal simulation is performed in an identical manner to the diffuse-sky simulation described in Section~\ref{sec:diffuse_foreground_sim}, but with the GSM healpix map replaced with a healpix map of uniform values corresponding to the value of the 21-cm model at that frequency.

\subsubsection{Internal noise coupling, beam mutual coupling and calibration errors}
\label{coupling}
 
It is an inherent property of closely-spaced interferometers, such as the EDA-2, that internal noise from the receiver electronics will be radiated by an antenna and subsequently received by neighbouring antennas, causing some correlation of internal noise at the correlator \citep{Venumadhav2016,ung2020,sutinjo2020}. This internal-noise coupling has its greatest effect at short baselines and will cause the interferometric response to change from the ideal case shown in Fig.~\ref{fig:1}. Since, as discussed in Section~\ref{sec:eda2_response_uniform}, it is the shortest baselines where the response to a uniform sky is at its greatest, this coupling is a concern to any experiment aiming to detect the global 21-cm signal with an interferometer. These effects require an in-depth treatment and we therefore leave their consideration to future work, which will examine the detectability of the global 21-cm in the presence of internal noise coupling and explore mitigation strategies to be employed in both system design and data analysis.

Another effect that has not been taken into account in this work is the mutual coupling between antenna elements that results in the antenna beam shapes deviating from idealised analytical models. Future work will incorporate detailed FEKO\footnote{https://altairhyperworks.com/product/Feko} models of the antenna beams for generating the expected uniform-sky response and creating simulated visibilities.

Real data will also never be calibrated perfectly and errors in the antenna-based phase and amplitude solutions will be present. We leave the inclusion of calibration errors in the simulations to future work as we continue to explore different calibration schemes and assess their accuracy.

\subsection{Real data from the EDA-2}
\label{sec:eda2-data-method}

\subsubsection{EDA-2 observations}

Observations were made with the EDA-2 on UTC~2020~March~3. The observations were carried out sequentially in frequency-channel order starting with channel 64 (centre frequency 50.0~MHz) at UTC~13:37:33 and finishing with channel 126 (centre frequency 98.9~MHz) at UTC~14:07:07. At each channel, 7 observations were made, each of duration 0.28~s. The observations made directly after and before changing channels were discarded due to known issues with the system at these `edge' times, so 5 observations for each centre frequency were used. The data at each coarse band were further channelised into 32 fine channels of width 28.935~kHz. The data were correlated and the resulting visibilities converted to measurement sets.

\subsubsection{EDA-2 calibration}
\label{sec:eda2_cal}

Each measurement set is calibrated using a full sky model as follows. We generate a GSM healpix map at the centre frequency of the coarse channel (as described in Section~\ref{sec:diffuse_foreground_sim}). We then make a dirty image from the uncalibrated measurement set using \sc{wsclean} \rm \citep{wsclean} to be used as a template. We use \sc reproject \rm and the header of the template image to create a reprojected map, which, following conversion from brightness temperature units to Jy/beam, is suitable as an input model image. The model image is predicted into the MODEL column of the measurement set using the \sc{wsclean} \rm option `-predict'. We then use \sc{calibrate}\rm, developed by \citet{offringa2016} for use on MWA data, to solve for the complex antenna gains, using this input model. We find that, due to signal-to-noise considerations for such a small amount of data, it is necessary in this first level of calibration to use all 32 fine channels to solve for a single frequency-independent solution per antenna. 

Once each measurement set in a coarse channel has been calibrated separately, all 5 are concatenated into a single measurement set using the \sc{casa} \rm \citep{casa} task \sc{concat} \rm. A second round of calibration is then performed on the CORRECTED data column of the concatenated data set using \sc{calibrate}\rm, but this time solving for the antenna gains on a per-fine-channel basis. The increased signal-to-noise ratio of the concatenated data set and the fact that a frequency-independent calibration has already been performed on the data, allows us to solve for and apply calibration solutions for each of the 32 fine channels and calibrate out any instrumental frequency structure occurring within a coarse band. The calibrated, concatenated data at each coarse channel is then exported to \sc{uvfits} \rm \citep{uvfits} format for further analysis.

\subsection{Signal extraction}
\label{sec:signal_extraction}

For both simulated and real data we read in the visibilities in \sc{uvfits} \rm format at each frequency. We extract the $(u,v)$ values from the 'UU' and 'VV' columns, convert them to units of wavelength and compute the baseline length as $\sqrt{u^{2}+v^{2}}$ for each pair. We then sort the $(u,v)$ values and their corresponding real visibility values in ascending order according to baseline length. A cut is made at a baseline length of $0.5\lambda$ and only data below this value are used in the subsequent analysis, since this is where there is a significant response to a global signal (see Fig.~\ref{fig:1}). The visibility values are then converted to brightness temperature units (K) by multiplying them by the conversion factor: $10^{-26}\lambda^{2}/2k$ (where $\lambda$ is wavelength in m and $k$ is the Boltzmann constant in m$^{2}$ kg s$^{-2}$ K$^{-1}$).

\begin{figure*}
  \centering 
  \subfloat[]{\includegraphics[clip,trim=5 5 30 40,width=0.5\textwidth]{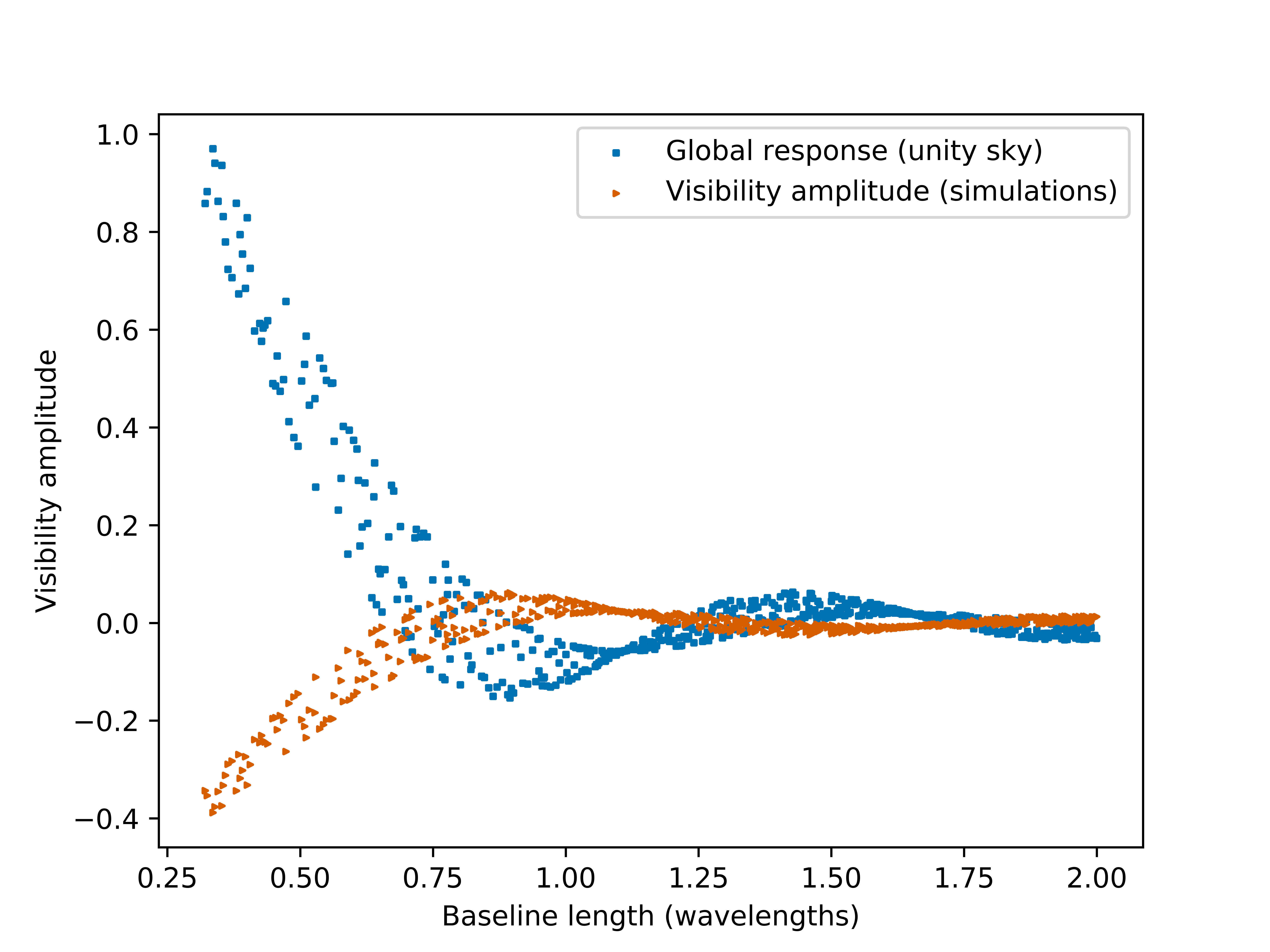}\label{fig:2a}}
  \hfill
  \subfloat[]{\includegraphics[clip,trim=0 5 30 40,width=0.5\textwidth]{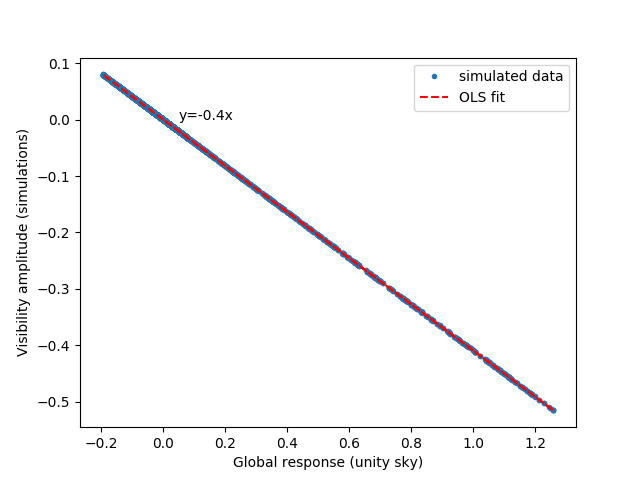}\label{fig:2b}}
  \caption{Simulation results for the E-W polarisation of an EDA-2 antenna layout at 70~MHz for simulation with a global redshifted 21-cm signal only as an input. (a): Expected response to a unity-valued uniform sky, normalised to 1 at the shortest baseline (blue squares) and the simulated visibility amplitude (orange triangles), vs baseline length for baselines shorter than 2$\lambda$ (only every 10th point is shown for clarity). The scaling factor between the expected uniform-sky response and the simulated visibility amplitude represents the global sky temperature, which is negative at this frequency for a foreground-free simulation (see Fig.~\ref{fig:3}). (b): Simulated visibility amplitude vs the expected response to a unity-valued uniform sky (blue points). The red dashed line is an OLS fit to the data, the gradient of which represents the measured global-sky temperature in K. This plot corresponds to the data in Fig.~\ref{fig:2a}.}
  \label{fig:2}
\end{figure*}

For each baseline we compute the expected response to a unity-valued, uniform sky according to equation~\ref{eqn:vis} (with $T_{\rm{sky}}(\nu)$ set to 1), using the analytic EDA-2 antenna beam shape described in Section~\ref{sec:eda2_response_uniform} in healpix form (in which case the integral can be evaluated as a sum over the pixel indices, since the pixels are of equal area on the sky). These values are the `global response' in equation~\ref{eqn:vis_angular}. As discussed in Section~\ref{sec:background_single_baseline}, the measured global sky temperature, $T_{\rm{sky}}(\nu)$ is the scaling factor that, when multiplied by the expected global response to a unity sky, gives the value of the visibility. This is shown in Fig.~\ref{fig:2a}, where we plot the expected global response (blue squares) and the visibility amplitude (orange triangles) as a function of baseline length at 70~MHz (for a global-21-cm-signal-only simulation). Both sets of points have been normalised so that the expected global response at the shortest baseline is unity. The measured global sky temperature, $T_{\rm{sky}}(\nu)$, is obtained by plotting the visibility amplitudes against the expected global responses, and fitting a line using a simple ordinary least squares (OLS) fit with a fixed intercept at the origin. An example plot at 70~MHz, which corresponds to the data in Fig.~\ref{fig:2a}, is shown in Fig.~\ref{fig:2b}.

This procedure is carried out at each frequency (each fine channel for the EDA-2 data and at 1~MHz intervals for the simulated data), resulting in a spectrum from the lowest frequency at 50~MHz up to the frequency where the number of baselines of length less than 0.5 wavelengths reaches zero. To assess the detectability of the global redshifted 21-cm signal in both the simulated and real data we use one of two simple metrics: the rms residual remaining after a) fitting and subtracting a polynomial to the data in log-log space, or b) jointly fitting a global redshifted 21-cm model and a log-log polynomial foreground model to the data. Following parts of \citep{bowman2018}, we use a simple polynomial foreground model, $T_{\rm{F}}(\nu)$, described by:
\begin{equation} 
T_{\rm{F}}(\nu)=\sum_{n=0}^{N-1}a_{n}\nu^{n-2.5},
\label{eqn:foreground}
\end{equation}
where N is the polynomial order, $a_{n}$ are the coefficients fitted to the data, $\nu$ is frequency and the factor of -2.5 in the exponent is used to match the expected spectral shape of the foreground. These are simplistic methods that could be unreliable for realistic global-signal detection \citep{gmoss2}, however, they are sufficient for the purposes of this work where the main focus is on demonstrating the validity of our techniques for using an interferometer to observe a global sky signal.

\subsubsection{Angular-response subtraction}
\label{sec:angular_response_removal}
As discussed in Section~\ref{sec:background_single_baseline}, the actual sky has angular structure and we would expect this to affect our signal extraction, where the visibility amplitudes are compared to the uniform-sky response. Anticipating the effect of angular structure, we compute the expected response to angular structure only, at each frequency, using the zero-mean sky map as described in Section~\ref{sec:diffuse_angular_foreground_sim}. When we employ the angular-response mitigation step in our data analysis we simply subtract the expected angular response from the measured (or simulated) visibility response before the OLS-fitting step described in Section~\ref{sec:signal_extraction} above. 

\section{Results}
\label{sec:results}

\subsection{Simulation results}

In this section we present results from simulations of the E-W polarisation of an EDA-2 antenna layout from 50 to 200~MHz with an integration time of 4~mins per 1~MHz channel. Synthetic data were generated following the methods outlined in Section~\ref{sec:method_sims} and the global sky signal was extracted following the method described in Section~\ref{sec:signal_extraction}.

\subsubsection{Simulation: Global redshifted 21-cm signal only}

\begin{figure}
  \centering
  \includegraphics[clip,trim=10 5 0 40,scale=0.5]{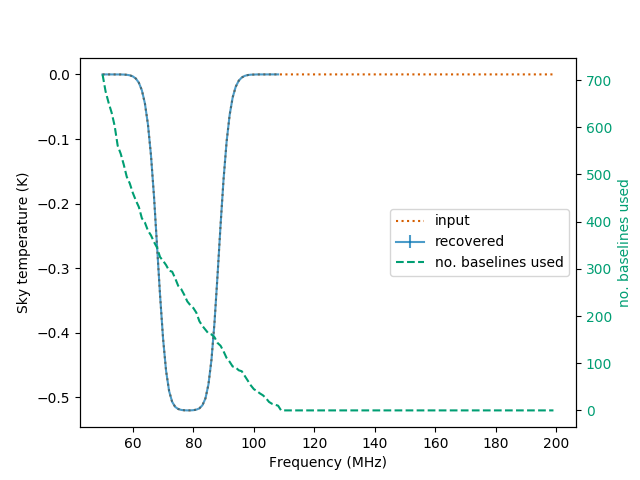}
  \caption{Simulation results for the E-W polarisation of an EDA-2 antenna layout for a simulation with a global redshifted 21-cm signal only as an input, across a frequency range 50-200~MHz. Orange dotted line: Input global 21-cm signal \citep{bowman2018}. Blue solid line: Recovered signal from our analysis pipeline. Green dashed line: number of baselines used in the analysis (those shorter than 0.5$\lambda$). The input signal is recovered perfectly up to approximately 110~MHz, at which point the number of baselines shorter than 0.5$\lambda$ reaches zero.}
  \label{fig:3}
\end{figure}

In Fig.~\ref{fig:3} we show the input and recovered signals for a simulation that contains only the global redshifted 21-cm signal (no noise) as an input. Fig.~\ref{fig:3} also shows the number of baselines used in the signal extraction (those shorter than 0.5$\lambda$) as a function of frequency. The plots in Fig.~\ref{fig:2} correspond to the point in Fig.~\ref{fig:3} at 70~MHz. The recovered signal is only extracted up to approximately 110~MHz, at which point the number of baselines shorter than 0.5$\lambda$ reaches zero.

\subsubsection{Simulation: Global redshifted 21-cm signal plus thermal noise}

\begin{figure*}
  \centering 
  \subfloat[]{\includegraphics[clip,trim=5 5 30 40,width=0.5\textwidth]{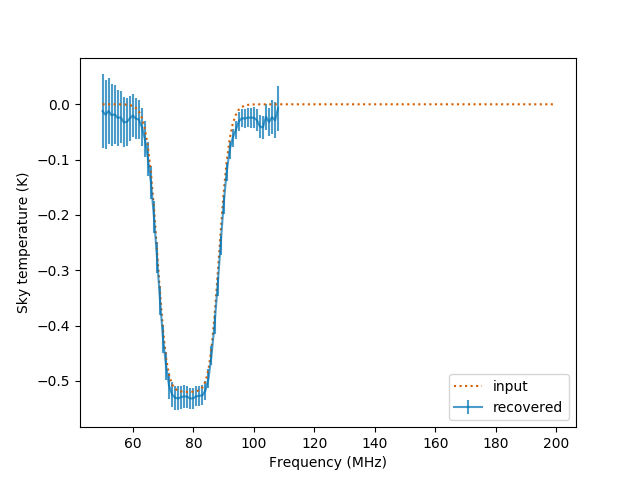}\label{fig:4a}}
  \hfill
  \subfloat[]{\includegraphics[clip,trim=0 5 30 40,width=0.5\textwidth]{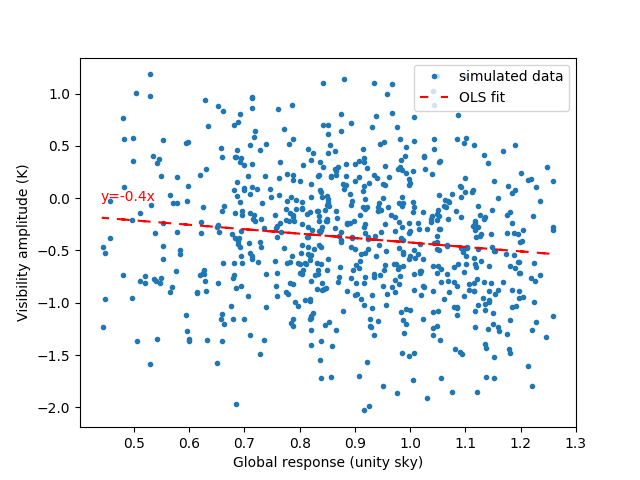}\label{fig:4b}}
  \caption{Simulation results for a 4-min simulated observation with the E-W polarisation of an EDA-2 antenna layout with a global redshifted 21-cm signal plus thermal noise as an input, across a frequency range 50-200~MHz. (a): Orange dotted line: input global 21-cm signal \citep{bowman2018}. Blue solid line with error bars: recovered signal from our analysis pipeline, using only baselines shorter than 0.5$\lambda$. The error bars are the standard errors from the OLS fits. The input signal is recovered up to approximately 110~MHz, at which point the number of baselines shorter than 0.5$\lambda$ reaches zero. (b): An example plot at 70~MHz of the simulated visibility amplitude vs the expected response to a unity-valued uniform sky (blue points). The red dashed line is an OLS fit to the data, the gradient of which represents the measured global-sky temperature in K. These data correspond to the brightness temperature value in Fig.~\ref{fig:4a} at 70 MHz.}
  \label{fig:4}
\end{figure*}

The results for a simulation including the global redshifted 21-cm signal plus thermal noise for a 4-min EDA-2 observation are shown in Fig.~\ref{fig:4}. Fig.~\ref{fig:4a} shows the input global redshifted 21-cm signal \citep{bowman2018} as an orange dotted line and the signal recovered with our analysis pipeline (using only baselines shorter than 0.5$\lambda$) as a blue solid line with error bars. The error bars are the standard error from the OLS fit at each frequency. Fig.~\ref{fig:4b} shows the simulated visibility amplitude vs the expected response to a unity-valued uniform sky (blue points) at 70~MHz. The OLS fit to the data is shown as a red dashed line, the gradient of which represents the measured global-sky temperature in K.

\subsubsection{Simulation: Global foreground signal plus thermal noise}
\label{sec:diffuse_global_noise_results}

\begin{figure*}
  \centering 
  \subfloat[]{\includegraphics[clip,trim=5 5 30 40,width=0.5\textwidth]{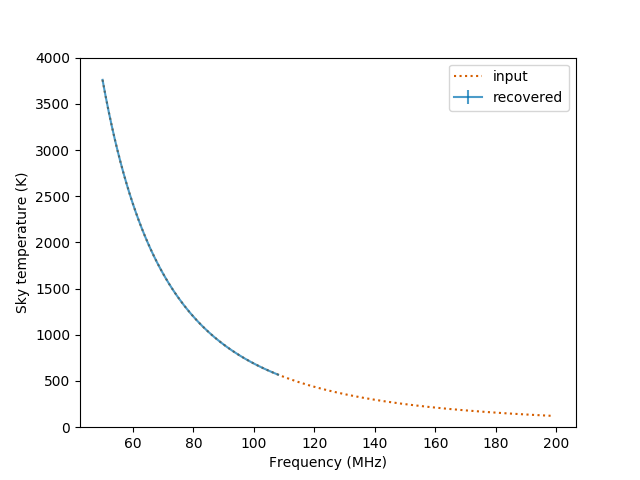}\label{fig:5a}}
  \hfill
  \subfloat[]{\includegraphics[clip,trim=0 5 30 40,width=0.5\textwidth]{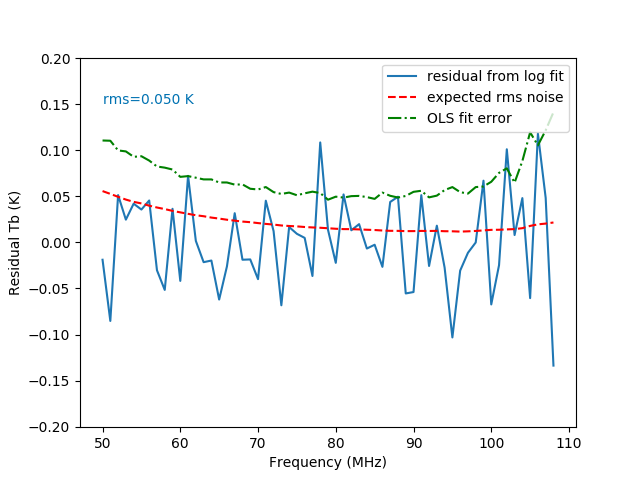}\label{fig:5b}}
  \caption{Simulation results for a 4-min simulated observation at LST~2~hrs with the E-W polarisation of an EDA-2 antenna layout with a diffuse global foreground signal (as described in Section~\ref{sec:diffuse_global_foreground_sim}) plus thermal noise as an input, across a frequency range 50-200~MHz. (a): Orange dotted line: input global foreground signal. Blue solid line with error bars: recovered signal from our analysis pipeline, using only baselines shorter than 0.5$\lambda$. The error bars are the standard errors from the OLS fits and are too small to be seen here, but their values are plotted in Fig~\ref{fig:5b} for comparison with the log-polynomial subtraction residuals. The input signal is recovered up to approximately 110~MHz, at which point the number of baselines shorter than 0.5$\lambda$ reaches zero. (b): Residuals after subtraction of a 5-th order polynomial in log-log space from the simulated data shown in Fig.~\ref{fig:5a} (blue solid line). The rms of the residuals is 50~mK. The expected thermal rms noise, calculated according to equation~\ref{eqn:thermal_noise} and taking into account the number of baselines included at each frequency channel, is shown as the red dashed line. The green dash-dot line is the OLS fit error at each frequency (corresponding to the small error bars in Fig~\ref{fig:5a}).}
  \label{fig:5}
\end{figure*}

In Fig.~\ref{fig:5} we show the results of a simulation that includes a foreground signal with no angular structure as described in Section~\ref{sec:diffuse_global_foreground_sim}, plus thermal noise. The LST used for the simulated observation is 2 hrs and no global redshifted 21-cm signal is included. In Fig.~\ref{fig:5b}
we show the residuals from the subtraction of a 5-th order polynomial fit in log-log space from the recovered signal (blue solid line) in Fig.~\ref{fig:5a}. The rms of the residuals is $\sim$~30~mK. Also shown in Fig.~\ref{fig:5b} is the expected rms thermal noise calculated according to equation~\ref{eqn:thermal_noise}. For the calculation, $T_{\rm{sys}}$ is taken to be the global-foreground temperature (orange dotted line in Fig.~\ref{fig:5a}), since sky noise dominates the system noise at these low frequencies. The effective area of the dipoles, $A_{\rm{eff}}$, is interpolated from the values calculated by \citet{wayth2017} for the EDA (which had an identical antenna layout to the EDA-2), and presented in their table~2. We take the values given in column~3 of their table~2 and divide by the number of antennas, 256. This properly takes into account the fact that the dipoles are embedded in a dense array. The calculation uses the number of baselines shorter than 0.5$\lambda$ at each frequency channel as $N$. 

\subsubsection{Simulation: Global foreground and redshifted 21-cm signal plus thermal noise}
\label{sec:foreground_plus_global}

\begin{figure*}
  \centering 
  \subfloat[]{\includegraphics[clip,trim=5 5 30 40,width=0.5\textwidth]{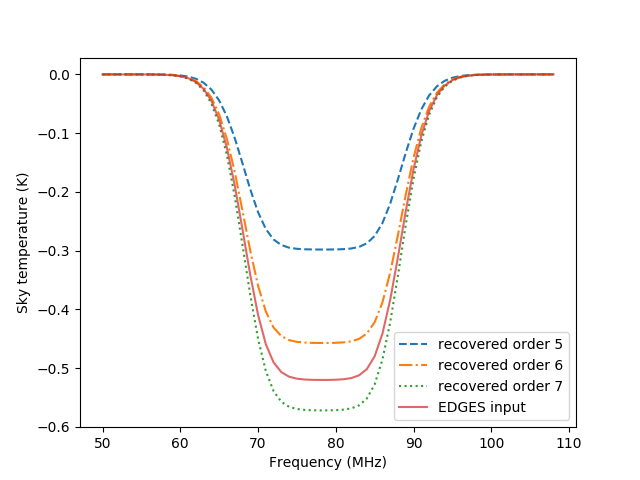}\label{fig:6a}}
  \hfill
  \subfloat[]{\includegraphics[clip,trim=0 5 30 40,width=0.5\textwidth]{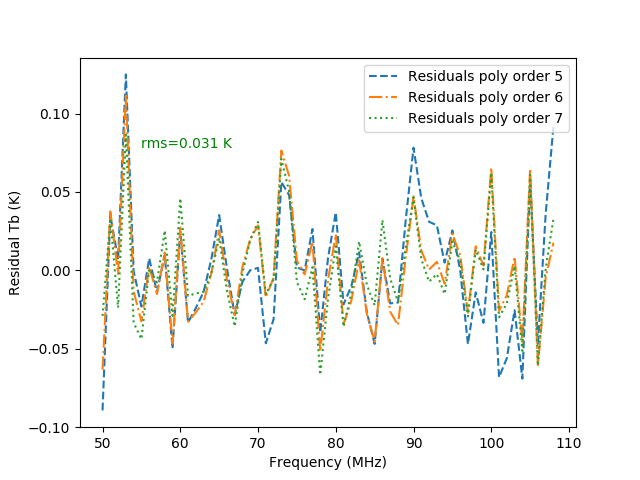}\label{fig:6b}}
  \caption{Simulation results for a 4-min simulated observation at LST~2~hrs with the E-W polarisation of an EDA-2 antenna layout with a diffuse global foreground signal, plus thermal noise and a global redshifted 21-signal as inputs (as described in Section~\ref{sec:foreground_plus_global}), across a frequency range 50-200~MHz. (a): Recovered global 21-cm signals from simulation and EDGES signal input (solid red line). The three recovered signals are from jointly fitting the EDGES model with a foreground model log-log polynomial of order 5 (blue dashed), 6 (orange dash-dot) and 7 (green dotted). (b): Residuals after subtraction of the 21-cm model and foreground polynomial for orders 5 (blue dashed), 6 (orange dash-dot) and 7 (green dotted). The rms value of 0.031~K shown refers to the 7th-order polynomial fit.}
  \label{fig:6}
\end{figure*}

In Fig.~\ref{fig:6} we show the results from a simulation with the same parameters as described in Section~\ref{sec:diffuse_global_noise_results}, but with the addition of a redshifted 21-cm signal. Fig.~\ref{fig:6a} shows the recovered global redshifted 21-cm signals from jointly fitting an EDGES model of the 21-cm signal and log-log foreground polynomials of orders 5, 6 and 7. The residuals after subtraction of the fitted foreground and 21-cm models are shown in Fig.~\ref{fig:6b}. The rms residual is approximately 30~mJy for all three polynomial orders.

\subsubsection{Simulation: Full sky model foreground and thermal noise}
\label{sec:results_full_sky}

\begin{figure*}
  \centering 
  \subfloat[]{\includegraphics[clip,trim=5 5 30 40,width=0.5\textwidth]{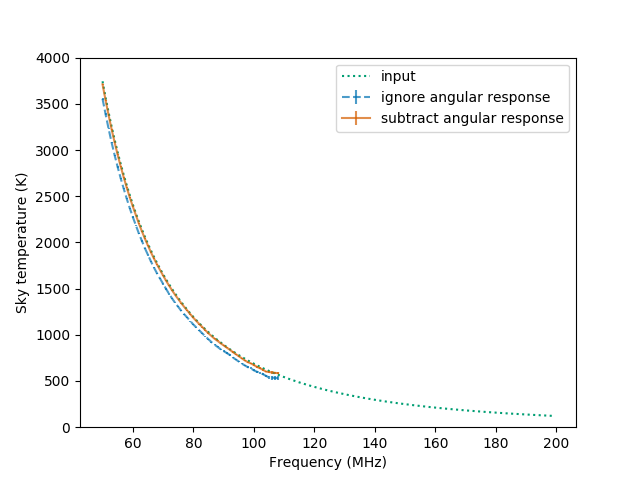}\label{fig:7a}}
  \hfill
  \subfloat[]{\includegraphics[clip,trim=0 5 30 40,width=0.5\textwidth]{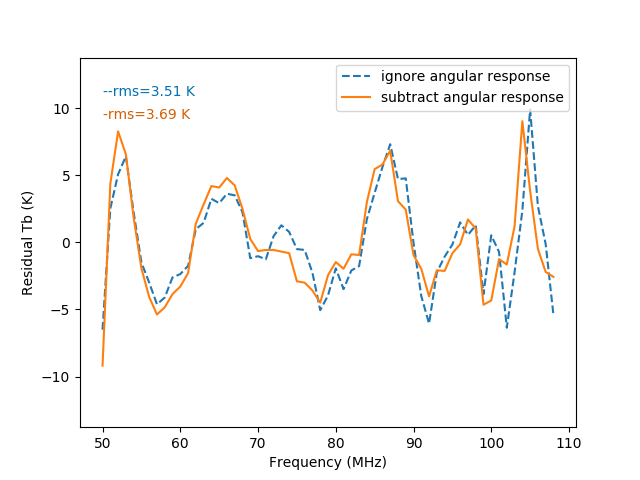}\label{fig:7b}}
  \caption{Simulation results for a 4-min simulated observation at LST~2~hrs with the E-W polarisation of an EDA-2 antenna layout with a full-sky-model foreground, plus thermal noise (as described in Section~\ref{sec:results_full_sky}), across a frequency range 50-200~MHz.(a): Input signal (green dotted line), recovered signal ignoring angular response (blue dashed line) and recovered signal using angular-response subtraction (orange solid line). (b): Residuals after subtraction of a 7-th order polynomial from the recovered signals shown in Fig.~\ref{fig:7a}. Blue dashed line: ignoring angular response, orange solid line: angular-response subtracted.}
  \label{fig:7}
\end{figure*}

In Fig.~\ref{fig:7} we show the results from a simulation including a full-sky-model foreground, as described in Section~\ref{sec:diffuse_foreground_sim}, plus thermal noise (no global redshifted 21-cm signal). The input and recovered signals are shown in Fig.~\ref{fig:7a} for the case where we ignore the interferometric response to the angular structure of the sky (in blue) and the case where we remove the expected angular response following the procedure outlined in Section~\ref{sec:angular_response_removal} (in orange). Fig.~\ref{fig:7b} shows the residuals after subtraction of a 7-th order log-log polynomial for both cases. Both sets of residuals have a similar shape, indicating that the angular structure introduces a systematic bias that is not removed by the angular-response mitigation process. The rms of both sets of residuals is above 3~K.

\subsection{Real data results: EDA-2}
\label{sec:eda2_results}

The EDA-2 data were observed and processed as detailed in Section~\ref{sec:eda2-data-method} and the global sky signal was extracted following the method described in Section~\ref{sec:signal_extraction}. 

\begin{figure*}
  \centering 
  \subfloat[]{\includegraphics[clip,trim=15 15 15 15,width=0.5\textwidth]{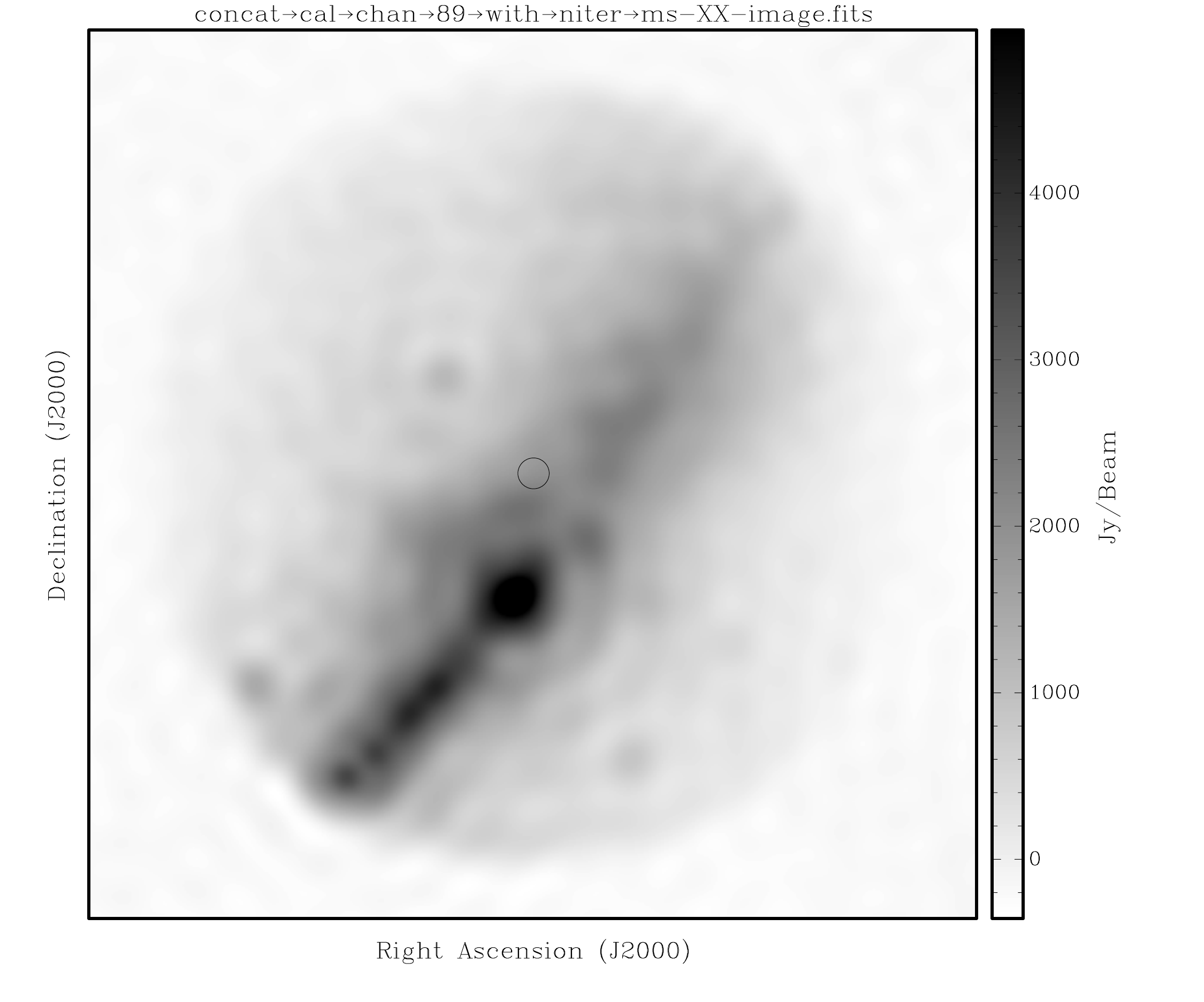}\label{fig:8a}}
  \hfill
  \subfloat[]{\includegraphics[clip,trim=0 0 45 40,width=0.5\textwidth]{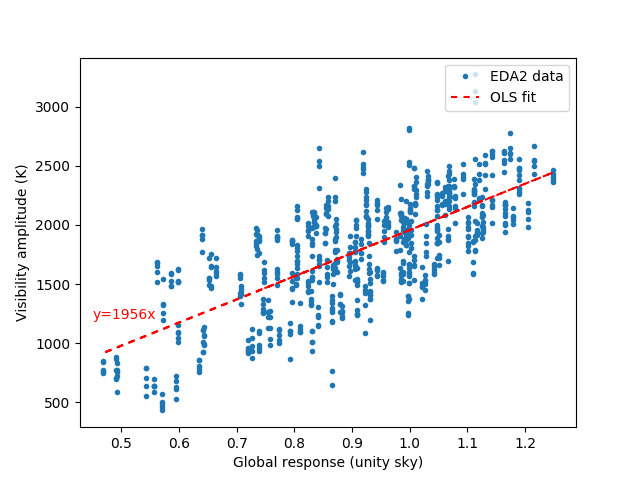}\label{fig:8b}}
  \caption{Real-data results from the EDA-2 at 70~MHz and LST 8.3~hrs. (a): CLEANed full sky image. The synthesis beam shape is shown at the centre of the image (zenith) (b): Visibility amplitude vs the expected unity-valued global-signal response (blue points) for the data corresponding to the image in Fig.~\ref{fig:8a}. The red dashed line is an OLS fit to the data, the gradient of which represents the measured global sky temperature in K.}
  \label{fig:8}
\end{figure*}

In Fig.~\ref{fig:8a} we show an example all-sky image using the full coarse band of data centred at 70~MHz (EDA-2 channel 89). The data were observed at LST 8.3~ hrs, so the image centre (zenith) is RA~(J2000)~8\textsuperscript{h}20\textsuperscript{m}, DEC $-26.7\degr$. The data have been CLEANed using \sc{wsclean} \rm and the synthesised beam, which has a major axis of $4.5\degr$, is shown in the centre of the image. In Fig.~\ref{fig:8b} we show the corresponding plot of the real component of the visibility amplitude vs the expected unity-valued global-signal response (blue points) at 70~MHz. The red dashed line is an OLS fit to the data, the gradient of which represents the measured global-sky temperature in K.

Initially, plots showing the real visibilities plotted against the expected unity-valued uniform-sky response (as in Fig.~\ref{fig:8b}) showed that there were outliers in the data in some coarse channels that were most likely caused by RFI. Automated RFI flagging was not conducted at any stage of the data processing due to the very short lever-arm available in both time and frequency, which makes identifying RFI by means of lines \citep{offringa2010}, or other morphologies \citep{offringa2012}, in time-frequency space very difficult. So, to excise these bad data, we took the linear model obtained from the OLS fit at each fine channel, subtracted it from the data and computed the mean and standard deviation of the residual. Any data points that were more than 5 standard deviations from the mean were removed from the data and the OLS fitting re-run on the `flagged' dataset. This removed most of the RFI-affected data. Very bad data are also excised by the fact that the calibration solutions do not converge, this results in some gaps in the measured spectrum where no observations at a particular frequency could be calibrated.

\begin{figure*}
  \centering 
  \subfloat[]{\includegraphics[clip,trim=0 0 40 30,width=0.5\textwidth]{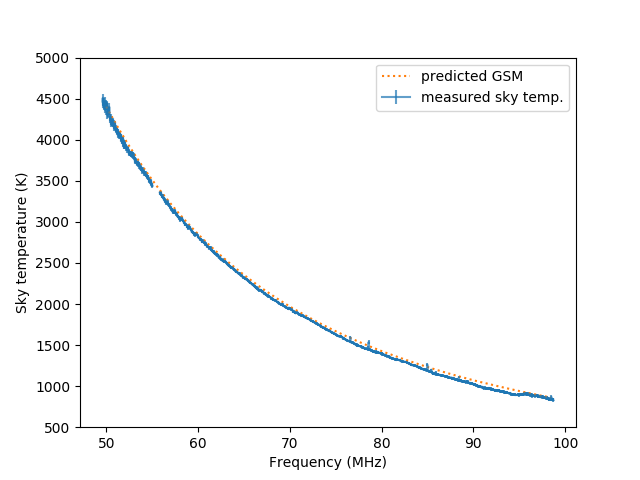}\label{fig:9a}}
  \hfill
  \subfloat[]{\includegraphics[clip,trim=0 0 40 30,width=0.5\textwidth]{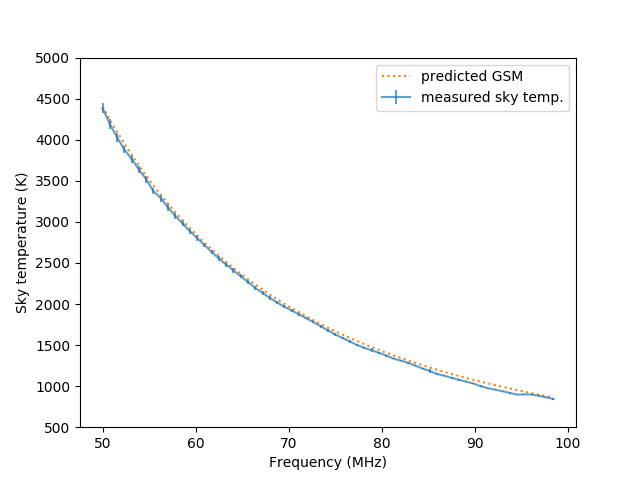}\label{fig:9b}}
  \caption{Real-data results from the EDA-2 at LST 8.3~hrs covering the frequency range 50-100~MHz. (a): Measured global sky temperature per fine channel of $\sim$~28~kHz (blue solid line and error bars) and expected global sky temperature derived from the GSM (orange dotted line). (b): Measured global sky temperature averaged per coarse channel of $\sim$~0.78~MHz (blue solid line and error bars) and expected global sky temperature derived from the GSM (orange dotted line).}
  \label{fig:9}
\end{figure*}

In Fig.~\ref{fig:9} we show the measured sky spectrum at both the full spectral resolution of 28~kHz (omitting the first 2 and final 3 fine channels in each of the overlapping coarse channels) in Fig.~\ref{fig:9a}, and with the data averaged within each coarse band in Fig.~\ref{fig:9b}, again using just the `central' 27 channels. The error bars in Fig.~\ref{fig:9a} are the standard error from the OLS fit at each fine frequency channel, while the error bars in Fig.~\ref{fig:9b} are the standard deviation of the measured sky temperatures from the individual fine channels used to calculate the per-channel average at each coarse band.

\begin{figure*}
  \centering 
  \subfloat[]{\includegraphics[clip,trim=0 0 40 30,width=0.5\textwidth]{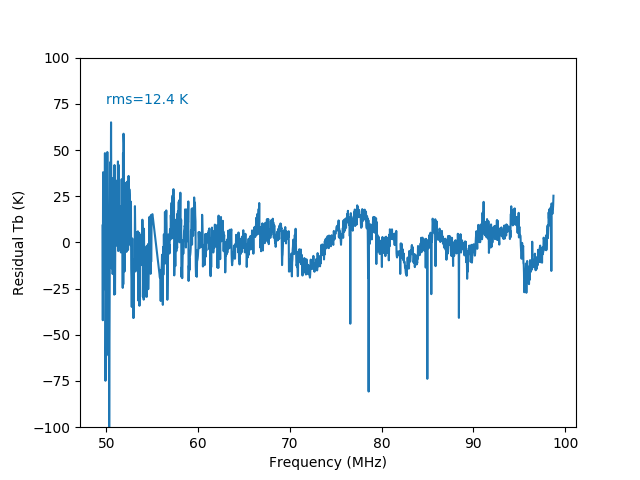}\label{fig:10a}}
  \hfill
  \subfloat[]{\includegraphics[clip,trim=0 0 40 30,width=0.5\textwidth]{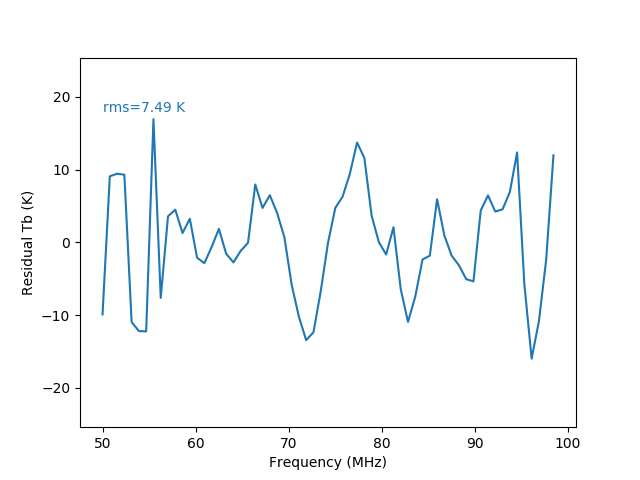}\label{fig:10b}}
  \caption{Real-data results from the EDA-2 at LST 8.3~hrs covering the frequency range 50-100~MHz. (a): residuals from subtracting 7th-order log-log polynomial from the fine-channel-resolution extracted signal shown in Fig.~\ref{fig:9a}. (b): residuals from subtracting 7th-order log-log polynomial from the coarse-channel-resolution extracted signal shown in Fig.~\ref{fig:9b}.}
  \label{fig:10}
\end{figure*}

In Fig.~\ref{fig:10} we show the residuals after subtracting a 7th-order polynomial from the measured spectrum, for both the full spectral resolution (Fig.~\ref{fig:10a}) and the  per-coarse-channel-averaged spectra (Fig.~\ref{fig:10b}).

\section{Discussion}
\label{sec:discussion}

The results of our analyses of both the simulated and real short-spacing interferometer data demonstrate significant progress toward measuring a global redshifted 21-cm signal interferometrically. Here we discuss the implications of these results with a view to planning future work.

\subsection{Discussion of simulation results}
\label{sec:sim_discussion}

The simplest simulation, that of a global redshifted 21-cm signal only, is a useful sanity check to show that both our simulations and signal-extraction tools are set up correctly. As expected and shown in Fig.~\ref{fig:3}, the input signal is recovered perfectly up until the point where the number of short baselines (length less than $0.5\lambda$) reaches zero at approximately 110~MHz. Similarly, for a global 21-cm signal with thermal noise, as shown in Fig.~\ref{fig:4a}, our analysis is able to recover the signal reasonably well, despite the large spread in the points contributing to the OLS fit (Fig.~\ref{fig:4b}). It is evident from Fig.~\ref{fig:4a}, however, that in the presence of noise the recovered sky temperature is lower than the input signal across the frequency range and the signal-extraction accuracy is reduced as the frequency approaches 110~MHz, where there are fewer short baselines available for the OLS fits. While these simulations without a foreground signal are unrealistic, they demonstrate that the signal-extraction pipeline works (even for noisy data) and serve as useful pedagogical examples to explain our newly-developed techniques. 

The recovered sky signal shown in Fig.~\ref{fig:5} for a diffuse-global foreground input signal (i.e. the GSM model averaged across the sky so that it has no angular structure, as described in Section~\ref{sec:diffuse_global_foreground_sim}) with thermal noise, but no global redshifted 21-cm signal, appears to match very well with the simulation input signal. Fig.~\ref{fig:5a} shows that the input signal is recovered accurately all the way from 50 to 110~MHz, at which point the number of baselines of length less than 0.5$\lambda$ reaches zero. 

In Fig.~\ref{fig:5b}, the 50~mK rms of the residuals (blue solid line) and the standard error of the OLS fits at each frequency (green dash-dot line) are greater than the expected rms noise as calculated from equation~\ref{eqn:thermal_noise} (red dashed line) for most of the band. This may be due to the limited wide-field accuracy of \sc{miriad}\rm, which is adding extra variation away from the unity-sky response into the simulated data used in the fits. A more detailed error analysis of the OLS fitting will be investigated in future work when it is planned to use a more accurate wide-field simulation tool.

Due to the large magnitude of the foreground signal the thermal noise has little effect on the results. An rms residual of 50~mK indicates that in the non-physical case of a uniform foreground signal, or assuming that the effects of angular structure can be fully removed, it would be possible to recover a global redshifted 21-cm signal of similar magnitude to the EDGES signal in a 4-min EDA-2 observation. This is also assuming a perfectly-calibrated, ideal instrument where the effects of coupling between interferometer elements (antennas and amplifiers) has been completely mitigated. Future work to include these effects in the simulations is discussed in Section~\ref{sec:future}.

When an EDGES-like global 21-cm signal is added to the simulation input with a uniform foreground signal and noise, we are able to perform a joint fit and recover the input 21-cm signal, as shown in Fig.~\ref{fig:6}. The results in Fig.~\ref{fig:6a} show that at least a 6th-order polynomial is required to fully recover the signal. The residuals after subtraction of the fitted foreground and 21-cm model (Fig.~\ref{fig:6b}) have an rms of approximately 30~mJy for all three polynomial orders, confirming that a global EDGES signal could be detected for an ideal instrument in the case where the foreground angular structure has been completely removed.

For the most realistic simulation, where we include a full GSM sky model with angular structure and thermal noise (but no global 21-cm signal), the signal-extraction accuracy is reduced. Fig.~\ref{fig:7a} shows that if we simply ignore the angular structure in the foreground (blue dashed line), the recovered sky temperature is lower than the input signal by between 1-4 percent, depending upon the frequency. This effect is similar to the result obtained for a global 21-cm signal with noise as shown in Fig.~\ref{fig:4}, indicating that angular structure in the sky has a similar effect to thermal noise on our estimates of the sky temperature for short snapshot observations (4 mins in this simulated case). 

Unlike the case of thermal noise, however, increasing the integration time will not necessarily decrease the bias introduced by angular structure. Only combining observations at sufficiently separated LSTs, where the interferometric angular response is different enough to integrate down, will help to increase the accuracy of the signal extraction. A more comprehensive set of simulations covering a wide range of LSTs will be required to determine how much separation in time is required for the correlation of the angular fluctuation response between snapshot observations to be sufficiently reduced. For the shortest baselines and longest wavelengths, the fringe pattern covers a large patch of sky and the response will not change significantly even over the course of an entire night. Fortunately, the baselines in this regime have only a low sensitivity to angular structure (and a corresponding large sensitivity to the global signal). For longer baselines and shorter wavelengths, the angular structure has a greater impact, but decorrelates faster. For example, the fringe pattern for an E-W baseline of length 1$\lambda$ has a width of approximately 3.8 hrs in RA, so the angular structure response will be completely decorrelated for observations spaced by a few hours in LST. An exploration of the optimal observing cadence, however, is left for future work.

In the presence of angular structure, the recovered signal (Fig.~\ref{fig:7a}, blue dashed line) is close to the input signal, but is not smooth as a function of frequency, resulting in much larger residuals when a polynomial foreground model is subtracted than the global foreground case, as shown in Fig.~\ref{fig:7b} (blue dashed line). These residuals due to the angular-structure response, with an rms of around 3.5~K, are more than two orders of magnitude greater than the residuals caused by thermal noise alone (e.g. Figs.~\ref{fig:5b} and \ref{fig:6b}). Thus, thermal noise can be considered insignificant compared to the impact of the angular structure response on the measured global sky signal for a snapshot observation.

When we remove the angular-structure response from the simulated data as described in Section~\ref{sec:angular_response_removal}, we recover the input global sky temperature more accurately (see Fig.~\ref{fig:7a}, orange solid line). However, this subtraction process does not mitigate against the unwanted spectral fluctuations that are evident in the residuals plotted in Fig.~\ref{fig:7b}. This is an unexpected result, as in simulation the angular response subtraction should be perfect, apart from the small effects of the thermal noise. The most likely explanation is that the widefield response to the sky simulated using \sc{miriad }\rm is not accurate to the required level, due to approximations made in the interest of simulation speed. Therefore, in future work a simulation package that is properly qualified for wide fields of view will be used.

With rms residuals of approximately 3.5~K for both the angular-response-subtracted and angular-response-ignored cases, it would not be possible to detect a global redshifted 21-cm signal in these simulated data and hence we do not proceed with adding a global 21-cm signal to the simulations, but leave this for future work when we combine multiple LSTs and explore more advanced methods of angular-response mitigation. This is discussed further in Section~\ref{sec:future}. 

\subsection{Discussion of real data from the EDA-2} 
\label{sec:eda2_discussion}

The all-sky image from the EDA-2 at 70~MHz (Fig.~\ref{fig:8a}) shows an accurate depiction of the expected sky at LST 8.3 hrs, with the Galactic plane stretching across the sky and the Gum nebula, including the bright Vela supernova remnant, a prominent feature close to zenith. This indicates that the in-field calibration as described in Section~\ref{sec:eda2_cal} has performed reasonably well. Observing when the Galactic plane was high in the sky was necessary for in-field calibration with a full sky model as it has been found that there is insufficient signal-to-noise for this calibration scheme to work for LSTs where the sky is more `empty'. Unfortunately, this is also the worst time to observe from the perspective of spectral structure being imprinted on the extracted global signal by Galactic foregrounds \citep{gmoss2}. To avoid this unwanted effect, it is possible to calibrate on either the Sun during the day, or the Galactic plane at suitable LSTs, and transfer solutions to observations taken at LSTs that are better-suited to global-signal measurements. This is not ideal, however, as the instrumental response is likely to change over the intervening time, due to temperature fluctuations and other effects. Reasonable images have been made with the EDA-2 from transferred calibration solutions, however, more work is required to test whether it is more advantageous to avoid Galactic structure by transferring calibration solutions or to deal with the angular and spectral structure of the Galactic plane in other ways. 

Even with in-field calibration using the bright Galactic plane, evidence that the calibration is imperfect can be seen in Fig.~\ref{fig:8b}, where the spread of points does not look noise-like as was seen in the simulations (see Fig.~\ref{fig:4b} for comparison). The banded structure apparent in Fig.~\ref{fig:8b} is likely to result from the same calibration solutions being applied to different observations, which were concatenated together to increase the signal-to-noise ratio for calibration purposes (as described in Section~\ref{sec:eda2_cal}). Groups of five points that appear spread in a vertical line at a specific x-axis value in Fig.~\ref{fig:8b} should have almost identical amplitudes as they are measuring the same sky modes only seconds apart. This effect could be mitigated by system improvements to allow longer observation times (including a wider instantaneous bandwidth so that the observer does not need to sequentially scan across frequencies), removing the need for the concatenation of shorter observations.

Despite the imperfect calibration of the instrument, the global-sky-signal measurements in Fig.~\ref{fig:9} show that the signal extraction has recovered the expected sky temperature reasonably well. At the full spectral resolution of 28~kHz (Fig.~\ref{fig:9a}) there are some obvious RFI spikes that have eluded our rudimentary flagging procedure (see Section~\ref{sec:eda2_results}). These RFI-affected fine channels are also clearly seen in the residual plot in Fig.~\ref{fig:10a}, and there is likely to be more RFI at lower levels. For the coarse-band-averaged spectrum in Fig.~\ref{fig:9b} (and the corresponding residuals in Fig.~\ref{fig:10b}) these bad channels have been left in the data and hence will contribute to some of the remaining spectral structure. 

A more concerning effect that is clearly seen in the full-spectral-resolution residuals of Fig.~\ref{fig:10a} is the structure that appears with a periodicity corresponding to the coarse-channel bandwidth (0.78~MHz). We suspect that this is due to the need for an initial frequency-independent calibration step for each coarse band, which may result in some spectral structure being imprinted at this frequency interval. Averaging the measured spectrum on a per-coarse-channel basis largely removes this effect, leaving only the larger-frequency-scale fluctuations, as can be seen by comparing Figs.~\ref{fig:10a} and ~\ref{fig:10b}.

Across most of the band, the measured sky temperature is slightly lower than the modelled temperature (Fig.~\ref{fig:9}), as we might expect from our full-GSM-model simulations described in Sections~\ref{sec:diffuse_foreground_sim} and \ref{sec:results_full_sky}. The values obtained for the measured sky temperature for the real data, however, were not appreciably changed by attempts to apply the angular-response removal procedure described in Section~\ref{sec:angular_response_removal} (and hence not shown). This suggests that calibration error, rather than the response to angular structure, could be the dominant effect causing the discrepancy between measured and modelled values (since, as discussed in Section~\ref{sec:sim_discussion}, the thermal noise contribution is insignificant). We are unable to confirm this, however, because we were not able to validate our angular response mitigation procedure, due to the limitations of our simulations.

\begin{figure}
  \centering
  \includegraphics[clip,trim=5 5 30 30,scale=0.55]{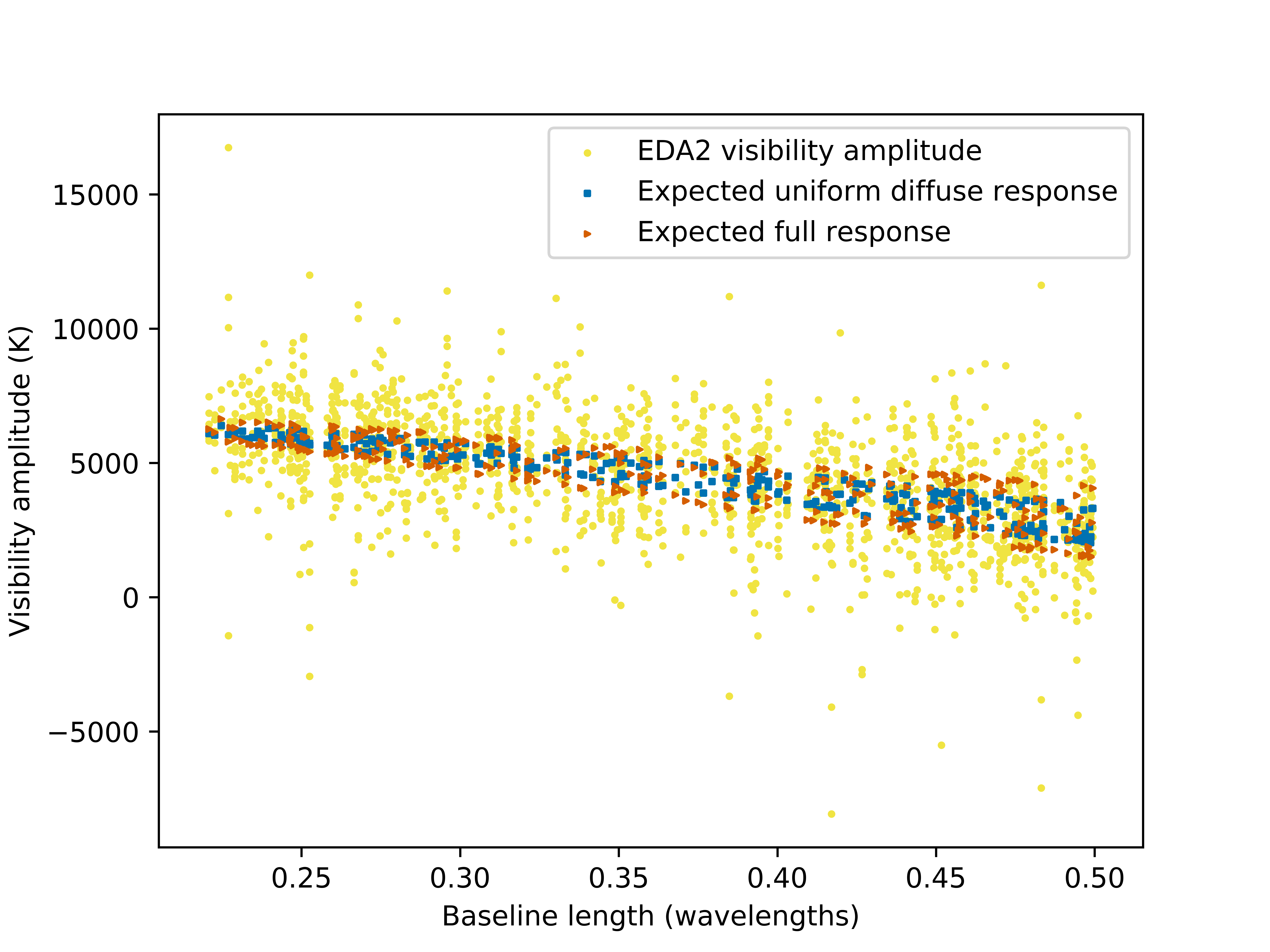}
  \caption{Visibility amplitude vs baseline length at 50~MHz for real EDA-2 measurements (yellow circles) and, calculated according to equation~\ref{eqn:vis_angular}, the expected `diffuse-global' foreground response (beam-weighted average of the GSM with no angular structure, blue squares), and the expected response to the full, spatially-varying GSM model (orange triangles). Note the greater variation in the measured data over the calculated full-GSM-model response.}
  \label{fig:11}
\end{figure}

To investigate the impact of angular structure and calibration errors further, we plot in Fig.~\ref{fig:11}, as a function of baseline length and for a frequency of 50~MHz, the measured visibility amplitudes for the EDA-2 observations (yellow circles) and the visibility amplitudes calculated according to equation~\ref{eqn:vis_angular} for both the `diffuse-global' foreground response (beam-weighted-average of the GSM with no angular structure) as blue squares and the expected response to the full GSM model as orange triangles. Fig.~\ref{fig:11} shows that the spread of the data away from the expected global-signal response is far greater for the measured data than for the full-GSM-model predicted response (calculated analytically). Some of the vertical banded structure of the measured points can be attributed to the calibration errors discussed above (and seen in Fig.~\ref{fig:8b}). We cannot rule out, however, the possibility that angular fluctuations are the dominant effect and that we are not modelling them correctly due to inaccuracies in either the sky or beam models used, or a combination of the two.

The rms variation of the measured spectrum, after averaging over each coarse channel (Fig.~\ref{fig:10b}), is approximately 7.5~K over the 50~MHz bandwidth used. While this is still at least 2 orders of magnitude worse than the level of spectral smoothness required for a global 21-cm detection, it is a promising result that is only a factor of $\sim2 $ larger than the rms residuals in our full-GSM model simulations (see Fig.~\ref{fig:7b}), which were simulated at a much more favourable LST of 2~hrs. While there are clearly unaccounted-for systematic errors affecting the shape of Fig.~\ref{fig:10b}, the results do show that we have achieved our initial aim to demonstrate that a global foreground signal can be extracted from a closely-spaced interferometer such as the EDA-2.

\section{Future work}
\label{sec:future}

Throughout this paper we have identified a number of areas for immediate future work. Based on the lessons learned, we lay out here a road map for both the simulations and real-data analysis that will ultimately lead to the development of a new, purpose-built instrument for the measurement of the redshifted global 21-cm signal; ASSASSIN.

\subsection{Future simulation work} 
\label{sec:future_sims}
Based on the results for a full-sky-model simulation (see Fig.~\ref{fig:7}), the next immediate steps for the simulation work must be to reduce the residual spectral structure that remains after subtraction of a foreground model to an acceptable level for 21-cm-signal detection. This work has identified that \sc{miriad }\rm is insufficient for future work as it does not accurately simulate wide fields of view to the level required for global 21-cm signal experimentation. Alternative simulation packages will be assessed for functionality and speed and the most suitable will be adopted for future work. Options include \sc{pyuvsim }\rm \citep{pyuvsim}, \sc{prisim }\rm \citep{prisim}, and \sc{woden }\rm \citep{woden}. 

The next set of simulations will be run across a wide range of LSTs and the optimal observational cadence for decorrelating the angular fluctuation response will be determined. They will also make use of both instrumental polarisations observed by the crossed-dipole antennas (this work took advantage of only half the data by examining just the E-W polarisation). These simulations will inform the development of the observing strategy for the next round of data collection. Once it is shown that a sufficiently smooth spectral response can be achieved with an ideal instrument, we will move on to introducing more complex instrumental effects.

Our results from real data show that calibration errors may have had a large impact on our measurements of the global sky signal (see Fig.~\ref{fig:8b}), however, in this work we cannot rule out that the dominant error is due to fluctuating foregrounds. The behaviour of the calibration errors encountered here are probably specific to the unusual, but necessary, calibration strategy used due to the way in which the data were acquired. The inclusion of calibration errors into future simulations will therefore be guided by the observing strategy that is to be adopted going forward. If through simulations we can identify an acceptable level of calibration error, this will help to inform both the development of immediate improvements to the EDA-2 and designs for ASSASSIN. 

As our observations become better-calibrated and more sensitive to the global sky signal, we are likely to start seeing systematic effects due to mutual coupling of both the antenna elements and their associated amplifiers. Hence, it is imperative that these effects are incorporated into the simulations. The first step will be to replace the analytic description of the antenna beam pattern by more accurate FEKO models that take into account mutual coupling. Such embedded element patterns (EEPs) have recently been generated by \citet{ung2020} for their work on characterising the noise temperature of the EDA in the presence of mutual coupling. We will investigate the use of individual EEPs for each dipole, as well as using a single, average element pattern, which may prove to be sufficient as was found by \citep{borg2020} for calibrating the Aperture Array Verification System-1 (Benthem et al., 2020, in preparation), albeit with less-stringent requirements for calibration accuracy than a global 21~cm signal detection will require.

Finally, internal noise coupling between array elements will affect the system noise temperature for an array of closely-spaced antennas \citep{Venumadhav2016,ung2019,ung2020,sutinjo2020}. Using the methods of \citet{ung2020}, we will be able to add in the effects of internal noise coupling to our simulations so that the simulated visibilities will include all known instrumental effects.

\subsection{Future EDA-2 observations}  

Until the design, development and deployment of ASSASSIN is complete, we will continue to make use of the EDA-2 for verification of signal-extraction techniques using real data and for experimentation with various calibration strategies. Guided by the outcomes of the simulation improvements described in Section~\ref{sec:future_sims}, the next round of observations will be conducted over a range of LSTs. We will also make use of data from both the E-W and N-S polarisations in our data analysis.

The observational results discussed in Section~\ref{sec:eda2_results}, indicate that calibration is possibly our biggest current limitation. To assess the true impact of calibration errors we must find a means to measure them independently of the angular structure response (i.e. not using the rms residual temperature). Possible ways to achieve this would be to look at the variation between observations that are closely spaced in time and therefore should have the same angular structure response, or to check the dynamic range of images made with the entire array.

The next set of observations will explore the strategy of transferring calibration solutions between pointings, rather that using in-field calibration, where we rely on the Galactic plane being high in the sky, potentially corrupting any global redshifted 21-cm signal. EDA-2 system upgrades may also allow us to change our current in-field calibration strategy. It is possible that a wider-band system could be deployed in the near future (possibly at the expense of the number of cross-correlated antennas), this would allow us to simultaneously observe a greater frequency range and remove the need for the multi-stage calibration procedure, which is likely responsible for the spectral structure we see within a coarse channel in our current data (see Fig.~\ref{fig:10a}). A wider instantaneous bandwidth would also allow us to perform more accurate RFI flagging. These system upgrades are currently under investigation. 

\subsection{Planning for ASSASSIN} 

We have used the existing EDA-2 to demonstrate that a global sky signal can be measured with a closely-spaced interferometer array, however, we anticipate that a new custom-built instrument will be required if we are to reach the level of precision required for a global-redshifted-21-cm-signal measurement. To this end, and using the lessons learned from the work described in this paper, we have begun planning for this new ASSASSIN array. The following work is either currently underway or will be commenced in the immediate future:
\begin{enumerate}[nosep,leftmargin=0.75cm]
    \item The development of a set of simplified design equations that can be used to include mutual-coupling effects into initial designs, without the need for more complex simulations and calculations \citep{sutinjo2020}.
    \item \label{point1} The designing, building and testing of a 2-element prototype short-spacing interferometer to verify our design equations and test performance in the field (Nambissan T. et al. 2020, in preparation).
    \item The design and implementation of a wideband backend to be initially tested on a subset of EDA-2 antennas.
    \item Development of robust observation and calibration strategies.
    \item Site planning and preparation for a complete, CD/EoR-science-ready ASSASSIN system.
\end{enumerate}
Future papers in this series will report on the progress of the ASSASSIN project.

\section{Conclusions}
\label{sec:conclusions}

In this paper we have used both simulated and real data to achieve three goals that represent significant progress toward measuring a 21-cm signal interferometrically with an array of closely-spaced dipoles. We have:
\begin{enumerate}[nosep,leftmargin=0.75cm]
    \item \label{point1} Outlined a new method for extracting a global sky signal from an array of closely-spaced antennas and used simulations to show that it is theoretically possible to measure a redshifted 21-cm signal in this way with an ideal instrument.  
    \item Verified our signal-extraction method using real data from the EDA-2 to measure the global Galactic foreground signal.
    \item Used the lessons learned in this work to lay out a plan for future work that will lead to the measurement of the redshifted 21-cm signal with ASSASSIN.
\end{enumerate}
Verifying the result of \citet{bowman2018} is a crucial next step to filling in the gap in our knowledge of the first billion years of the Universe's evolution. In particular, projects such as ASSASSIN, which seek to measure the global redshifted 21~cm using novel techniques, are important as they provide a means to avoid the same types of systematic errors that may be affecting the EDGES result. They can potentially offer both a completely independent verification of the signal at CD and provide a new means to explore later epochs such as the EoR.

\section*{Acknowledgements}

This scientific work makes use of the Murchison Radio-astronomy Observatory, operated by CSIRO. We acknowledge the Wajarri Yamatji people as the traditional owners of the Observatory site. We acknowledge that the Bentley campus of Curtin University, upon which most of this paper was written, is located on the land of the Noongar people. BM and CMT are supported by an Australian Research Council Future Fellowship under grant FT180100321. Some of the results in this paper have been derived using the \sc{healpy} \rm and \sc{healpix} \rm packages.

\section*{Data availability}
The data and code underlying this article will be shared on reasonable request to the corresponding author.

 



\begin{thebibliography}{99}
\bibitem[Astropy Collaboration(2013)]{astropy1}Astropy Collaboration, 2013, A\&A, 558, 33 
\bibitem[Astropy Collaboration(2018)]{astropy2}Astropy Collaboration, 2018, AJ, 156, 123
\bibitem[Barry et al.(2019)]{barry2019}Barry N., et al., 2019, in press, arXiv:1909.00561
\bibitem[Bernardi et al.(2015)]{leda2}Bernardi G., McQuinn M., Greenhill L. J., 2015, ApJ, 799, 90
\bibitem[Borg et al.(2020)]{borg2020}Borg J., Magro A., Zarb Adami K., de lera Acedo E., Sutinjo A., Ung D., 2020, MNRAS.tmp.1550B
\bibitem[Bowman et al.(2013)]{bowman2013}Bowman J. D., et al., 2013, PASA, 30, 31
\bibitem[Bowman et al.(2018)]{bowman2018}Bowman J. D., Rogers A. E. E., Monsalve R. A., Mozdzen T. J., Mahesh N., 2018, Nature, 555, 67
\bibitem[Bradley et al.(2019)]{bradley2019}Bradley R. F., Tauscher K., Rapetti D., Burns, J. O., 2019, ApJ, 874, 153
\bibitem[Burns et al.(2019)]{burns2019}Burns J. O., Bale S., Bradley R. F., 2019, American Astronomical Society Meeting \#234, id. 212.02. Bulletin of the American Astronomical Society, Vol. 51, No. 4
\bibitem[Burns et al.(2020)]{burns2020}Burns J. O., Hallinan G., 2020, American Astronomical Society meeting \#235, id. 130.01. Bulletin of the American Astronomical Society, Vol. 52, No. 1
\bibitem[Cornwell (1988)]{cornwell1988}Cornwell T. J., 1988, A\&A, 202, 316
\bibitem[Cranmer et al.(2017)]{lwa2}Cranmer M. D., Barsdell B. R., Price D. C., et al., 2017, Journal of Astronomical Instrumentation, 6, 1750007
\bibitem[Datta et al.(2016)]{datta2016}Datta, A., Bradley R., Burns, J. O., Harker, G., et al., 2016, ApJ, 831, 6
\bibitem[de Oliveira-Costa et al.(2008)]{gsm}de Oliveira-Costa A., Tegmark M., Gaensler B. M., Jonas J., Landecker T. L., Reich P., 2008, MNRAS, 388, 247
\bibitem[DiLullo et al.(2020)]{lwa3}DiLullo C., Taylor G. B., Dowell J., 2020, arXiv:2005.10669
\bibitem[Dowell et al.(2017)]{dowell2017}Dowell J., Taylor G. B., Schinzel F. K., Kassim N. E., Stovall K., 2017, MNRAS, 469, 4537
\bibitem[Ekers \& Rots(1979)]{ekers1979}Ekers R. D., Rots A. H., Image Formation from Coherence Functions in Astronomy, Proceedings of IAU Colloq. 49, held in Groningen, Netherlands, August 10-12, 1978. Edited by C. van Schooneveld, C. D. Reidel Publishing Co. (Astrophysics and Space Science Library. Volume 76), 1979., p.61
\bibitem[Furlanetto et al.(2006)]{furlanetto2006} Furlanetto, S. R., Oh, S. P., Briggs, F. H., 2006, PhR, 433, 181
\bibitem[G\'{o}rski et al.(2005)]{healpix}G\'{o}rski K. M., Hivon E., Banday A. J., Wandelt B. D., Hansen F. K., Reinecke M., Bartelmann M., 2005, ApJ, 622, 759
\bibitem[Greisen(2012)]{uvfits}Greisen, E. W. 2012, AIPS Memo Series, 117
\bibitem[Greenhill \& Bernardi(2012)]{leda1}Greenhill L. J., Bernardi G., 2012, HI Epoch of Reionization Arrays. Invited review to the 11th Asian-Pacific Regional IAU Meeting 2011, NARIT Conference Series, Vol. 1 eds. S. Komonjinda, Y. Kovalev, and D. Ruffolo, arXiv:1201.1700
\bibitem[Hills(2019)]{hills2019}Hills R., Kulkarni G., Meerburg P. D., Puchwein E., 2018, Nature, 564, 32
\bibitem[Jordan et al.(2017)]{jordan2017}Jordan C. H., et al., 2017, MNRAS, 471, 3974
\bibitem[Kraus(1988)]{kraus}Kraus J. D., 1988, Antennas, second edition, McGraw Hill, New York, USA
\bibitem[Lanman et al.(2019)]{pyuvsim}Lanman A. E., et al., 2019, Journal of Open Source Software, 4(37), 1234, https://doi.org/10.21105/joss.01234
\bibitem[Line et al.(2020)]{woden}Line J. L. B., Mitchell D. A., Pindor B., et al., 2020, PASA, 37, 27
\bibitem[Madau et al.(1997)]{madau1997}Madau P., Meiksin A. \& Rees M. J., 1997, ApJ, 475, 429
\bibitem[Mahesh et al.(2015)]{mahesh2015}Mahesh N., Subrahmanyan R., Udaya Shankar N., Raghunathan A., 2015, IEEE Transactions on Antennas and Propagation, vol. 63, isssue 11, pp 4835 - 4847
\bibitem[McKinley et al.(2013)]{mckinley2013}McKinley B., et al., 2013, AJ, 145, 23
\bibitem[McKinley et al.(2018)]{mckinley2018}McKinley B., et al., 2018, MNRAS, 481, 5034
\bibitem[McMullin et al.(2007)]{casa}McMullin, J., Waters, B., Schiebel, D., et al., 2007, in ADASS XVI, vol. 376 of ASPCS, 127
\bibitem[Mozdzen et al.(2017)]{mozdzen2017}Mozdzen T. J., Bowman J. D., Monsalve R. A., Rogers A. E. E., 2017, MNRAS, 464, 4995
\bibitem[Offringa et al.(2010)]{offringa2010}Offringa A. R., de Bruyn A. G., Biehl M., et al., 2010, MNRAS, 405, 155
\bibitem[Offringa et al.(2012)]{offringa2012}Offringa A. R., van de Gronde J. J., Roerdink J. B. T. M., 2012, A\&A, 539, 95
\bibitem[Offringa et al.(2014)]{wsclean}Offringa A. R., et al., 2014, MNRAS, 444, 606
\bibitem[Offringa et al.(2016)]{offringa2016}Offringa A. R., et al., 2016, MNRAS, 458, 1057
\bibitem[Presley, Liu \& Parsons(2015)]{presley2015}Presley, M. E., Liu A., Parsons A. R.
\bibitem[Price(2016)]{pygsm}Price D. C., 2016, Astrophysics Source Code Library, record ascl:1603.013
\bibitem[Rhodes(2011)]{pyephem}Rhodes B. C., 2011, Astrophysics Source Code Library, record ascl:1112.014
\bibitem[Sathyanarayana Rao et al.(2017a)]{gmoss}Sathyanarayana Rao M., Subrahmanyan R., Udaya Shankar N., Chluba J.,  2017, AJ, 153, 26
\bibitem[Sathyanarayana Rao et al.(2017b)]{gmoss2}Sathyanarayana Rao M., Subrahmanyan R., Udaya Shankar N., Chluba J., 2017, ApJ, 840, 33
\bibitem[Sault, Staveley-Smith \& Brouw(1996)]{sault1996}Sault R. J., Staveley-Smith L., Brouw W. N.,  1996, A\&AS, 120, 375
\bibitem[Sault, Teuben \& Wright(1995)]{miriad}Sault R. J., Teuben P. J., Wright M. C. H., Astronomical Data Analysis Software and Systems IV, ASP Conference Series, Vol. 77, 1995, R.A. Shaw, H.E. Payne, and J.J.E. Hayes, eds., p. 433.
\bibitem[Shaver et al.(1999)]{shaver1999}Shaver P. A., Windhorst R. A., Madau P., de Bruyn A. G., 1999, A\&A, 345, 380
\bibitem[Singh et al.(2015)]{singh2015}Singh S., Subrahmanyan, R., Udaya Shankar N., Raghunathan, A., 2015, ApJ, 815, 88
\bibitem[Singh et al.(2019)]{singh2019}Singh, S., Subrahmanyan R., 2019, ApJ, 880, 26
\bibitem[Sokolowski et al.(2015)]{sokolowski2015}Sokolowski, M., Wayth R. B., Tremblay, S. E., Tingay, S. J., et al., 2015, ApJ, 813, 18
\bibitem[Sutinjo et al.(2020)]{sutinjo2020}Sutinjo A. T., et al., 2020, accepted for publication in Union de Radio-Scientifique Internationale Radio Science Letters, arXiv:2009.04643
\bibitem[Taylor et al.(2012)]{lwa1}Taylor G., Ellingson S., Kassim N., et al., 2012, Journal of Astronomical Instrumentation, 1, 1250004
\bibitem[Thompson(1999)]{thompson}Thompson A. R., 1999, in Taylor G. B., Carilli C. L., Perley R. A., eds, ASP Conf. Ser. Vol. 180, A Collection of Lectures from the Sixth NRAO/NMIMT Synthesis Imaging Summer School, Astron. Soc. Pac., San Francisco, p. 11 
\bibitem[Thyagarajan et al.(2015)]{thyagarajan2015}Thyagarajan, N., et al., 2015, ApJ, 804, 14
\bibitem[Thyagarajan et al.(2020)]{prisim}Thyagarajan N., Harish S., Kolopanis M., et al., 2020, nithyanandan/PRISim v2.2.1 (Version v2.2.1), Zenodo, http://doi.org/10.5281/zenodo.3892099
\bibitem[Tingay et al.(2013)]{tingay2013}Tingay S. J., et al., 2013, PASA, 30, 7
\bibitem[Trott et al.(2018)]{trott2018}Trott C. M., et al., 2018, ApJ, 867, 15
\bibitem[Turner(2016)]{skalow}Turner W., 2016, Technical Report SKA-TEL-SKO-0000008, SKA Phase 1 System Requirements Specification (Cheshire:SKA Organisation)
\bibitem[Ung(2019)]{ung2019}Ung D. C. X., 2019, Masters of Philosophy Thesis, Curtin University, https://espace.curtin.edu.au/bitstream/handle/20.500.11937/77989/Ung\%20D\%202019.pdf
\bibitem[Ung et al.(2020)]{ung2020}Ung D. C. X., Sokolowski M., Sutinjo A. T., Davidson D. B., 2020, IEEE Trans. Antennas. Propagat., accepted, 6 Mar, 10 pages, 13, arXiv:2003.05116
\bibitem[van Haarlem et al.(2013)]{lofar}van Haarlem M. P., Wise M. W., Gunst, A. W., et al., 2013, A\&A, 556, 2
\bibitem[Vedantham et al.(2014)]{vedantham_iono}Vedantham H. K., Koopmans L. V. E., de Bruyn A. G., Wijnholds S. J., Ciardi B., Brentjens M. A., 2014, MNRAS, 437, 1056
\bibitem[Vedantham et al.(2015)]{vedantham2015}Vedantham H. K., et al., 2015, MNRAS, 450, 2291
\bibitem[Venumadhav et al.(2016)]{Venumadhav2016}Venumadhav T., Chang T-C., Dor\'{e} O., Hirata C. M., 2016, ApJ, 826, 116
\bibitem[Wayth et al.(2017)]{wayth2017}Wayth R. B., et. al., 2017, PASA, 34, 34
\bibitem[Wrobel \& Walker(1999)]{noise_eqn}Wrobel J. M., Walker R. C., 1999, in Taylor G. B., Carilli C. L., Perley R. A., eds, ASP Conf. Ser. Vol. 180, A Collection of Lectures from the Sixth NRAO/NMIMT Synthesis Imaging Summer School, Astron. Soc. Pac., San Francisco, p. 171
\bibitem[Zaldarriaga et al.(2004)]{zaldarriaga2004}Zaldarriaga M., Furlanetto S. R. \& Hernquist L., 2004, ApJ, 608, 622
\bibitem[Zheng et al.(2017)]{zheng2017}Zheng, H., et al., 2017, MNRAS, 464, 3486
\bibitem[Zonca et al.(2019)]{healpy}Zonca et al., 2019, Journal of Open Source Software, 4(35), 1298
\end{thebibliography}



\bsp	
\label{lastpage}
\end{document}